\documentclass[aps,prapplied,reprint,superscriptaddress,amsmath,amssymb,floatfix,nobalancelastpage]{revtex4-2}

\usepackage{graphicx}
\usepackage{bm}
\usepackage{xcolor}
\usepackage{xr}
\newcommand{\enx}{\varepsilon_{n,x}}
\newcommand{\eny}{\varepsilon_{n,y}}
\newcommand{\kick}{\mathrm{kick}}

\newcommand{\closingheading}[1]{\par\addvspace{2.2ex}\noindent{\large\bfseries #1}\par
  \nobreak\vspace{1.0ex}\noindent\ignorespaces}
\AtBeginDocument{%
  \renewenvironment{acknowledgments}{\closingheading{Acknowledgments}}{\par}}

\begin{document}

\title{Physics-constrained graph closures and working-point conditioning\\
for a fast surrogate of linac beam dynamics}

\author{Bowen Zhang}
\affiliation{Shanghai Advanced Research Institute, Chinese Academy of Sciences, Shanghai 201210, China}
\affiliation{University of Chinese Academy of Sciences, Beijing 100049, China}
\author{Dazhang Huang}
\email{huangdz@sari.ac.cn}
\affiliation{Shanghai Advanced Research Institute, Chinese Academy of Sciences, Shanghai 201210, China}
\author{Nengyuan Zhang}
\affiliation{Shanghai Advanced Research Institute, Chinese Academy of Sciences, Shanghai 201210, China}
\affiliation{University of Chinese Academy of Sciences, Beijing 100049, China}
\author{Duan Gu}
\author{Chao Feng}
\email{fengc@sari.ac.cn}
\author{Bo Liu}
\affiliation{Shanghai Advanced Research Institute, Chinese Academy of Sciences, Shanghai 201210, China}

\date{\today}

\begin{abstract}
When a graph neural network estimates a physical quantity, its frame, edge direction, aggregation
and feature normalization decide which laws it can break, and wrong choices leave systematic errors
the loss never registers. We build a linac beam-dynamics surrogate on this observation. Space charge,
short-range wakefields and coherent synchrotron radiation enter as bounded closures on analytic
kernels; the graphs carry a frame and a direction dictated by the interaction; and their symmetry
makes the net space-charge self-force vanish for any weights. Verified without training, with
randomized weights, the invariants break by five to eight orders of magnitude once the responsible
feature is inverted. Per-working-point models constrained by the rms envelope equation are distilled
into one model conditioned on 15 machine settings. On the 18.4~m L0+L1 section of the Shanghai soft
x-ray free-electron laser linac, this model predicts the transverse beam envelope at 50 unseen
settings to about $7\%$, together with the centroid trajectory and emittance, in 0.38~s on a single
Quadro RTX 6000 GPU. Training is instead dominated by a normalization choice, a data-derived output
range for the beam size: at all 16 working points it lowers the starting error 2.2- to 8.5-fold and
fixes the convergence speed. The remaining accuracy differences between hybrid and data-driven
architectures vary with the working point and with the density of observation. The held-out error
scales roughly as $N^{-0.6}$ with the corpus size $N$. Whether the physics pays off where collective
effects are strong, as in a magnetic bunch compressor, is in the next phase of our study.
\end{abstract}

\maketitle

\section{Introduction}
\label{sec:intro}

Machine learning (ML) is being pressed into service as a potential substitute for expensive physical
simulations, and three recipes for embedding physical laws have become standard. Physics-informed
neural networks add a governing-equation residual to the loss~\cite{raissi2019}. Structure-preserving
architectures build a symmetry or conservation law into the network itself: Hamiltonian and
Lagrangian networks do this for energy~\cite{greydanus2019,cranmer2020}, equivariant graph networks
for geometric symmetry~\cite{satorras2021}, and learned graph-network simulators~\cite{sanchezgonzalez2020}
extended with antisymmetric messages~\cite{sharma2026} for momentum. A third recipe puts the unknown
part of a model as a learned closure inside known equations~\cite{rackauckas2020}. Neural operators
belong to a related but distinct data-driven route: they learn solution maps of parametric equations
directly from data, with no residual, conservation or symmetry constraint imposed~\cite{li2021fno}.

A graph network that estimates a physical interaction, rather than classifying or regressing an
abstract target, faces a further set of choices that is rarely examined. Which frame defines the
neighbors? Are the edges directed? How are messages aggregated? Which inputs get normalized? Each
answer fixes which physical properties the network is able to violate. Several defaults of standard
graph networks turn out to be inappropriate for a Coulomb-like interaction inside a relativistic
particle bunch. They fail \emph{silently}: training converges, the loss looks unremarkable, and the
learned quantity is systematically biased whenever particle number or bunch density changes.

Daily operation and commissioning of an accelerating facility---in our case the linear accelerator
(linac) of a free-electron laser (FEL)---turn on adjusting magnetic field strengths, radio-frequency
(rf) phases and amplitudes, and other key parameters until the machine performs as required. Every
adjustment moves the machine to a new \emph{working point}. Before applying it to the real machine,
an operator wants to know what that working point does to the beam along the beamline: transverse
beam size, central orbit, emittance, energy, and so on. Tracking codes such as
\textsc{elegant}~\cite{borland2000,wang2006}, \textsc{astra}~\cite{floettmann2017astra} and
\textsc{impact-t}~\cite{qiang2000impact,qiang2006impactt} resolve the full dynamics (for \textsc{elegant}, see Sec.~\ref{sec:data}) with collective effects
included. On a common Linux server even the short section studied here costs them tens of seconds
with ten thousand macroparticles, and more than ten minutes with one million. Accelerator surrogates
nowadays do comparable work within a second or even much less, and yet gaps remain. Envelope models run in
milliseconds but treat collective effects through low-order closures. Accelerator
surrogates~\cite{edelen2024} have mostly followed the beam along the line at a fixed
setting~\cite{ivanov2020,garciacardona2024}, learned the map of a fixed section from its input beam
to its output beam~\cite{huang2023,sun2023}, or mapped settings to the output beam
only~\cite{edelen2020,gupta2021} without describing how the beam evolves. Networks trained on
collective effects~\cite{wan2025,leon2025} have generally left the underlying collective physics
unconstrained by construction.

Our contributions are the following.
(i)~\emph{Collective effects as bounded closures on analytic kernels, on graphs whose topology is set
by the physics.} Every message is an analytic kernel multiplied by a bounded, zero-initialized
modulation. The untrained model is therefore the analytic solution, and no input can move it by more
than a fixed factor. The graph is built in the frame where the interaction is electrostatic, is
directed only where the interaction is causal, and its message symmetry makes the net space-charge
self-force vanish identically (Sec.~\ref{sec:collective}).
(ii)~\emph{A catalog of silent failure modes} of standard graph-network defaults in this setting,
each removed by construction (Table~\ref{tab:silent}).
(iii)~\emph{Verification without training.} Because the invariants are structural, each is tested
with randomized weights against a design in which the responsible feature is inverted, on a
configuration constructed to make the invariant exact---in effect a unit test for an inductive bias
(Sec.~\ref{sec:checks}).
(iv)~\emph{Identifiability by construction of a second observable.} Two collective mechanisms
supervised by one quantity can be traded against each other at constant loss. An exact split of the
emittance growth into intraslice and slice-displacement channels gives a necessary condition for
telling them apart (Sec.~\ref{sec:bridge_main}).
(v)~\emph{Conditioning across machine settings.} Per-working-point models constrained by the envelope
equation are distilled into one model conditioned on the settings, after removing the dominant
failure mode of that step, the collapse of a micron-scale output onto the dataset mean
(Sec.~\ref{sec:generalization}).
(vi)~\emph{Separating physics from normalization.} An ablation over five seeds that equalizes the
data-driven baseline's output range shows that, in this test case, a data-derived output range for
the beam size affects convergence more strongly than any physics component at all 16 working points
tested, and accuracy more strongly with dense observation. A comparison between physics-informed and
data-driven models therefore speaks to the physics only once such choices are shared
(Sec.~\ref{sec:experiments}).
(vii)~\emph{The full evolution of the beam.} The envelope, centroid trajectory and emittance are
predicted and shown along the beamline at any working point, as a tracking code would show them, not
only at the output. At the optimal working point an additional head also predicts the full
six-dimensional distribution along the line, the longitudinal phase space included (Supplementary
Section~\ref{sm-app:scope}).

The model is evaluated on the $18.4$~m L0+L1 section of the linac of the Shanghai soft x-ray FEL user
facility (SXFEL-UF)~\cite{zhao2017}, with \textsc{Pelegant} tracking as ground truth
(Sec.~\ref{sec:data}), against a purely data-driven model. Ablation separates the contributions of
the physics components from those of the architecture (Sec.~\ref{sec:experiments}). Inputs are beam
data at observation points and the machine settings; no neural network (NN) model of any individual
magnet or cavity is needed, which can also be counted as an advantage of our approach. The intended
role is a fast complement to a tracking code, which stays the reference.

\section{Problem setting and data}
\label{sec:data}

\paragraph*{Background.} An electron bunch in a linac can be represented by a distribution of
macroparticles in $(x,x',y,y',t,\delta)$---transverse positions, emission angles, arrival time and
relative energy deviation. Operators watch its rms sizes $\sigma_{x,y}$ (the \emph{envelope}), its
centroid $\langle x\rangle,\langle y\rangle$ (the \emph{orbit}), its emittance $\varepsilon$ (the
phase-space area occupied), and similar quantities. Along a drift or through a magnet, the envelope
obeys the rms envelope equation, Eq.~(\ref{eq:kv}) below. \emph{Collective effects} are the forces
the bunch exerts on itself through inter-particle interactions and through the interaction of
particles with the structures. Three are considered here: (i) space charge, the Coulomb repulsion of
the particles; (ii) short-range wakefields, left in an accelerating structure by the head of the
bunch and acting on its tail; and (iii) coherent synchrotron radiation (CSR), emitted in bending
magnets by the tail and overtaking the head. \emph{Watch points} are the positions where the tracking
code records the bunch; on a real machine they correspond to beam diagnostics. Long-range wakefields,
which drive multi-bunch instabilities, are ignored here because the distance between bunches is much
larger than the bunch length---with a repetition rate of 10 to 50~Hz and a 10~ps bunch, the trailing
bunch never sees the wake of the one before it.

\begin{figure}[t]
\centering
\includegraphics[width=\columnwidth]{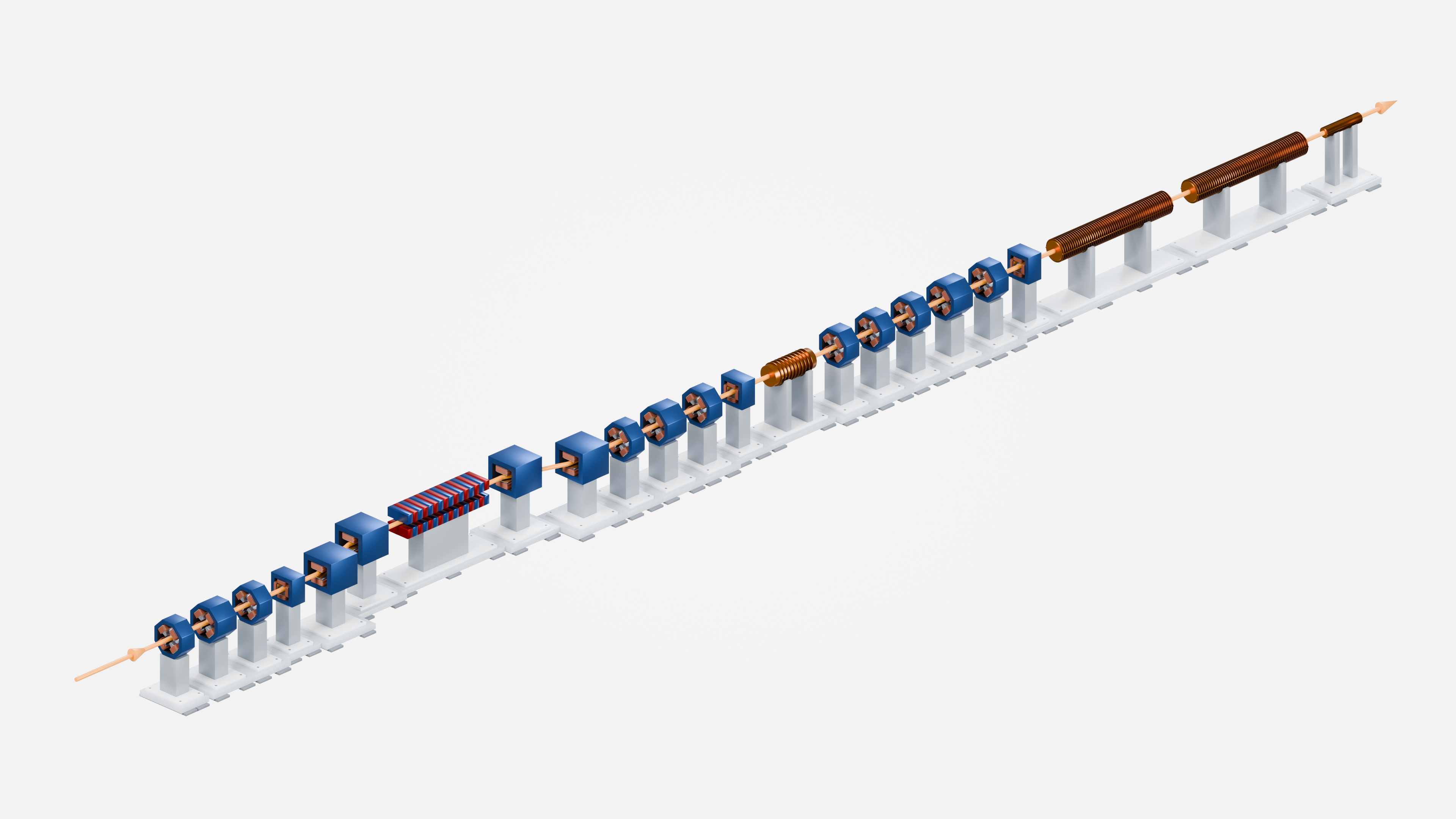}
\caption{Layout of the L0+L1 section of the SXFEL-UF linac used as the test case.}
\label{fig:sxfelinj}
\end{figure}

\paragraph*{Beamline and dataset.} The L0+L1 section of the SXFEL-UF linac (Fig.~\ref{fig:sxfelinj})
takes the beam from about $130$ to $252$~MeV at the usual operating point, and consists of a laser
heater (one magnetic chicane and one undulator), three quadrupole triplets, one FODO cell, three skew
quadrupoles, two S-band accelerating structures and one X-band linearizer. This part was chosen
because the triplets and the FODO cell form a dedicated beam-diagnostic section, able to host more
profile and position monitors than the rest of the linac, so the surrogate can later be compared with
measurements on the real machine. Ground truth is \textsc{Pelegant}
tracking~\cite{borland2000,wang2006} with longitudinal space charge, short-range wakefields and CSR,
recorded at 57 watch points. \textsc{elegant} was chosen because an accelerator lattice can be
designed and varied in it simply and flexibly; it is a standard tool for linac design and for the
microbunching studies that will be carried out in phase II of this
work~\cite{borland2008,huangsxfel2011}; and its tight integration with the SDDS
toolkit~\cite{borland1995sdds,borland1998toolkit} makes data acquisition and handling straightforward.
All of these features make \textsc{elegant} a convenient tool for charged-particle tracking.
\textsc{elegant} usually does not model transverse space charge in a linac, which is weak above
$130$~MeV as in our test case: over most of the section its defocusing strength, of order
$10^{-3}$~m$^{-2}$, lies about three orders of magnitude below that of the quadrupoles, and an
envelope calculation along the lattice puts its effect on the beam size at a few percent on average,
larger only near the beam waists and with peak current (Supplementary
Section~\ref{sm-app:data}). None of the comparisons in this work depends on it, since every model is
trained and tested on the same reference, and the collective module already contains a transverse
space-charge branch for a reference that includes it. More broadly, neither the aim nor the
conclusions of this work turn on it: the claims are methodological and relative---the invariants
verified without training, the catalogued silent failure modes, and the weight of physics against
normalization---so what is at stake is the surrogate's consistency with its ground truth, not that
reference's own fidelity to the real machine. The accuracy findings are in any case stated only for
this weakly collective regime, of which weak transverse space charge is one feature (discussed in
Sec.~\ref{sec:limits}). For more accurate predictions on the real machine, and to put the transverse
space-charge branch of the collective module to full use, which is exactly what we want eventually, the next stage will calibrate it against the real measurements and/or against tracking with full space-charge codes such as
\textsc{astra} and \textsc{impact-t}. 

A working point is the vector
\begin{equation}
w = \big(k_1^{(1)},\dots,k_1^{(11)};\ \varphi_1,\varphi_2;\ V_1,V_2\big)\in\mathbb{R}^{15},
\end{equation}
of eleven quadrupole strengths, two rf phases and two rf voltages. The laser-heater parameters are
fixed once the heater has been optimized, and are therefore excluded from $w$. The dataset holds 500
working points, sampled uniformly within $\pm20\%$ in each quadrupole strength, $\pm10^{\circ}$ in
each phase and $\pm10\%$ in each voltage around the optimal operating point of the real machine. 50
working points, \texttt{sxfel001}--\texttt{sxfel050}, are held out of training entirely, so the test
measures generalization to unseen settings. The conditioned model predicts transverse quantities;
the reason is discussed in Supplementary Section~\ref{sm-app:data}, together with the beam and
lattice parameters, the tracking setup, and the distribution of the watch points. All three
collective effects of the reference, the longitudinal short-range wake included, are also in the
collective module (Sec.~\ref{sec:collective}); their reach into the transverse quantities is sized
in Supplementary Section~\ref{sm-app:data}.
The longitudinal phase space itself is predicted by a separate head, so far at the optimal working
point: it maps the entrance macroparticles to any position $s$ through a polynomial map whose
coefficients a network computes as a function of $s$, and reproduces the six-dimensional
distribution, $t$ and $p$ included, down to the least-squares floor of a smooth single-particle map
(Supplementary Section~\ref{sm-app:scope} and Figs.~\ref{sm-fig:psexit} and \ref{sm-fig:psres}).
Longitudinal moments across working points---bunch length, energy spread and chirp---are not yet
outputs of the conditioned model; they become essential in a bunch compressor and are left to the
next phase (Sec.~\ref{sec:limits}). The distribution of the watch points
is uneven by construction of the optics---nearly a third of the watch points sit in a $1.7$~m stretch
around the beam waist, while the $7.7$~m occupied by the accelerating structures is observed only in
three short gaps.

\section{Method}
\label{sec:method}

\subsection{Surrogate overview}
\label{sec:model}

\begin{figure*}[t]
\centering
\includegraphics[width=0.78\textwidth]{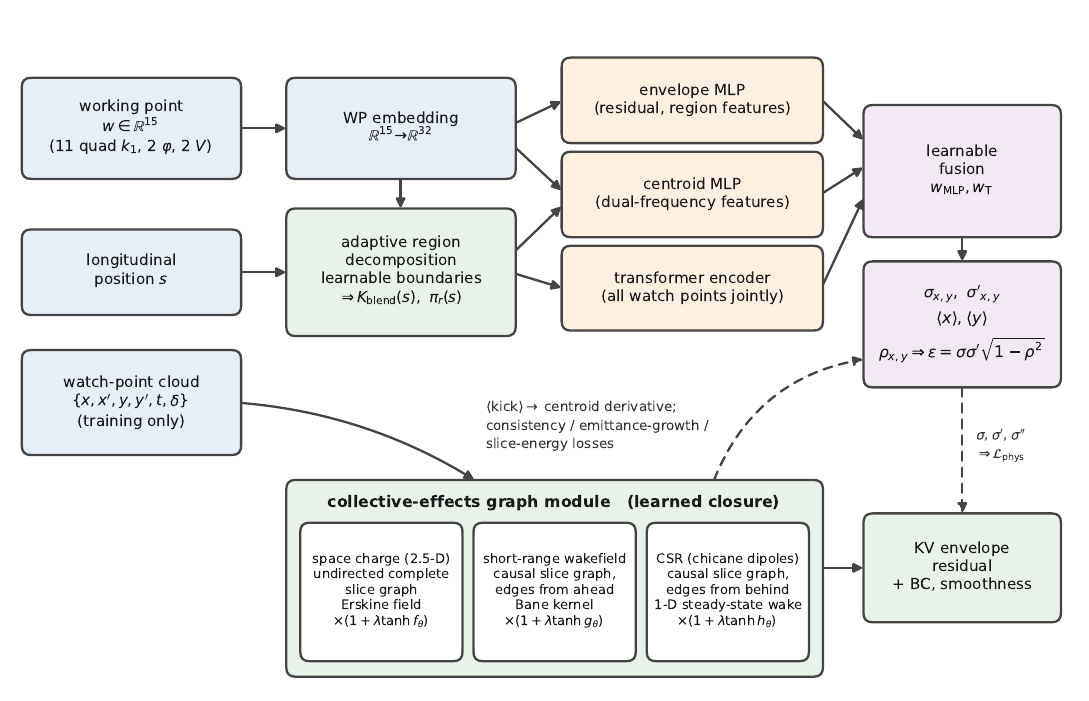}
\caption{Architecture. The working-point vector is embedded and injected into every prediction path.
A learnable decomposition of the beamline into longitudinal regions supplies the blended focusing
strength $K_{\rm blend}(s)$ and the soft region memberships $\pi_r(s)$ as input features. Envelope
and centroid are predicted by separate residual MLPs, in parallel with a transformer encoder that
sees the ordered sequence of all watch points; two learnable weights, normalized to sum to one,
combine the two paths. The collective module computes space charge, short-range wakefields and
coherent synchrotron radiation from analytic kernels modulated by a bounded graph closure, and enters
through the space-charge term of the envelope residual and through the bunch-averaged kick on the
centroid derivative. Dashed lines act during training only.}
\label{fig:arch}
\end{figure*}

Architecture is shown in Fig.~\ref{fig:arch}. The map is
$(w,s)\mapsto(\sigma_{x},\sigma_{x'},\sigma_{y},\sigma_{y'},\langle x\rangle,\langle
x'\rangle,\langle y\rangle,\langle y'\rangle,\rho_x,\rho_y)$, with $s$ the longitudinal position and
$\rho$ the position--angle correlation (the six-dimensional head of Sec.~\ref{sec:data} is a
separate map of the entrance macroparticles); the emittance follows from
\begin{equation}
\varepsilon = \sigma\,\sigma'\sqrt{1-\rho^{2}} .
\label{eq:emit}
\end{equation}
Particle distributions enter during training only; inference needs $w$, the watch-point coordinates
and the fixed entrance distribution, and nothing else.

Envelope and centroid obey different equations and are predicted by separate residual MLPs acting
pointwise in $s$. Focusing along a linac is piecewise, so the beamline is divided into $R=4$ regions
by soft boundaries with learnable positions and sharpnesses (the last position is fixed at $0.95$ of
the section length by the parameterization). The positions are parameterized as a softmax over gaps
followed by a cumulative sum, which makes their order automatic. Each region carries a small network
for the focusing strength, and their blend $K_{\rm blend}(s)$ enters the envelope features and the
physics residual. A two-layer transformer encoder~\cite{vaswani2017} sees the ordered watch-point
sequence, and two learnable weights, normalized to sum to one, fuse it with the pointwise branch.
Features, region parameterization and fusion are in Supplementary Section~\ref{sm-app:arch}.

In the per-working-point models the envelope is constrained by the rms envelope
equation~\cite{kapchinskij1959} with a space-charge term,
\begin{equation}
\sigma_x''+\frac{\gamma'}{\beta^{2}\gamma}\sigma_x'+K_x\sigma_x
-\frac{\enx^{2}}{\beta^{2}\gamma^{2}\sigma_x^{3}}
-\frac{K\,F_x}{2(\sigma_x+\sigma_y)}=R_x ,
\label{eq:kv}
\end{equation}
and its vertical counterpart, where $\gamma$ is the Lorentz factor, $\enx$ the normalized emittance,
$K=2I/(I_A\beta^{3}\gamma^{3})$ the generalized perveance with $I$ the beam current and
$I_A\simeq17$~kA, and $F_{x,y}$ a shape factor supplied by the collective module. Derivatives come
from automatic differentiation with respect to $s$, and the residual $R_x$ is evaluated at
collocation points drawn uniformly along the line~\cite{raissi2019}, so Eq.~(\ref{eq:kv}) constrains
the network between watch points as well as at them. It enters the loss through the soft saturation
$\langle R^{2}\rangle/(1+\langle R^{2}\rangle)$, together with a boundary condition at both ends, a
smoothness penalty between adjacent watch points and a midpoint curvature penalty.

\subsection{Collective effects as bounded closures on analytic kernels}
\label{sec:collective}

The collective module supplies forces rather than predicting $\sigma(s)$; it is a solver, not a
feature extractor. Every message has the form
\begin{equation}
m_{ij}=k_{\rm analytic}(r_{ij})\,\big[1+\lambda\tanh f_\theta(\cdot)\big],
\label{eq:closure}
\end{equation}
with $\lambda\le0.3$ and the last layer of $f_\theta$ initialized to zero. Scaling, singularity
structure and asymptotics live in the analytic kernel; the network expresses only what the idealized
kernel omits---non-Gaussian or hollow transverse profiles, halo, image charges, slice-to-slice
correlations. The modulation is identically zero before training, so the module \emph{is} the
analytic solution, and outside the training distribution its output stays within a factor
$1\pm\lambda$ of that solution instead of diverging. Kernels are the Bassetti--Erskine field of an
elliptical Gaussian slice for space charge~\cite{bassetti1980}, the Bane kernel for the short-range
wake~\cite{bane2003}, and the one-dimensional steady-state wake for CSR~\cite{saldin1997};
Supplementary Section~\ref{sm-app:collective} gives them with the implementation.

\begin{figure*}[t]
\centering
\includegraphics[width=0.84\textwidth]{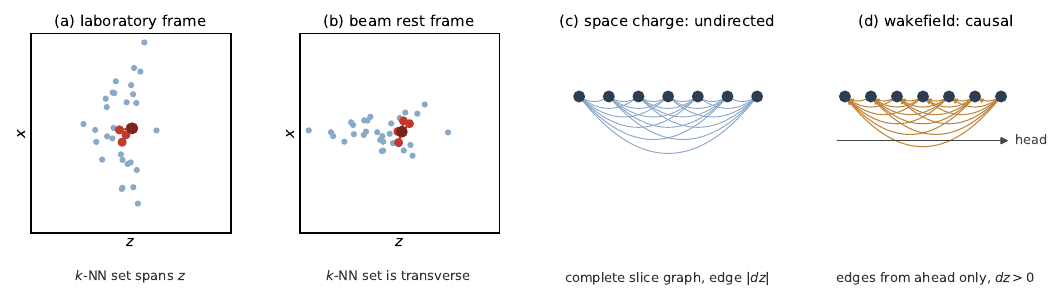}
\caption{Graph topologies are set by the interaction, not by tuning. (a),(b) The same
macroparticles in the laboratory frame and in the beam rest frame, where the longitudinal
coordinate is stretched by $\gamma$ (compressed here to $\gamma=6$ for legibility; the real value
rises from about 250 to 570 along the section). The four nearest neighbors of the marked particle,
in red, are almost disjoint between the two frames. (c) Space charge has no causal ordering, so the
slice graph is complete and undirected with $|dz|$ as the only edge feature. (d) A short-range wake
acts only backwards, so the wakefield graph is directed, with edges from slices ahead of the witness
slice.}
\label{fig:topo}
\end{figure*}

\paragraph*{Topology set by the interaction.} In the beam rest frame, where a relativistic bunch is
stretched by $\gamma$ and needle-like, space charge is electrostatic. A graph built on laboratory
coordinates therefore groups particles that are not neighbors in that frame; nearest-neighbor sets
from the two frames are almost disjoint [Fig.~\ref{fig:topo}(a,b)], and the graph is built on the
boosted coordinates. A pairwise sum in the rest frame, at the few thousand macroparticles a tracking
file holds, evaluates two-body collisions rather than a smooth self-field, so the default solver uses
longitudinal slices as nodes, which makes the result independent of the sampling. Space charge
carries no causal ordering, so the slice graph is complete and undirected [Fig.~\ref{fig:topo}(c)].
A short-range wake acts only backwards, so the wake graph is directed, with edges from slices ahead
of the witness [Fig.~\ref{fig:topo}(d)]. CSR is causal in the opposite sense.

\paragraph*{Conservation by construction.}
\label{sec:symmetry}
Momentum conservation of the space-charge interaction is a property of the architecture here, not an
approximate result of a penalty. A penalty holds inside the training distribution; an architectural
constraint on the conservation law holds for any input, whether or not the slice approximation is
accurate there. What enforces it is the symmetry of the tensor fed to the closure network. Pairwise
antisymmetry of the messages is an established route to momentum-conserving graph networks in other
physical systems~\cite{sharma2026}; here it is combined with analytic kernels
whose modulation is bounded, and with graphs whose frame and causal direction follow the interaction.
Closure inputs of the pair-sum space-charge graph, of the per-particle slice closure and of the wake
graph are
\begin{align}
e^{\rm sc}_{ij}&=\big[h_i+h_j,\ h_i\odot h_j,\ \log u,\nonumber\\
&\quad \tfrac{1}{u^{2}+1},\ \tfrac{u}{1+u},\ \log_{10}r_{\rm ref}\big],
\label{eq:edge_sc}\\
p^{\rm slice}&=\big[h_{\rm slice},\ u_x^{2},\ u_y^{2},\ u_x^{2}+u_y^{2}\big],
\label{eq:edge_slice}\\
e^{\rm wake}&=\big[h_{\rm wit},\ h_{\rm src},\ \tfrac{ds}{\mathrm{span}},\
   \sqrt{ds/s_0},\ \sqrt{ds/s_1}\big],
\label{eq:edge_wake}
\end{align}
where $h_i$ is the latent feature of node $i$, $u=r/r_{\rm ref}$ the dimensionless edge length and
$u_{x,y}=\Delta_{x,y}/\sigma_{x,y}$ the particle offset from its slice centroid. Equation
(\ref{eq:edge_sc}) contains permutation-invariant combinations only: the sum and elementwise product
of node features, plus scalar functions of the edge length. Messages of the slice graph use the same
symmetric combination $[h_i+h_j,\ h_i\odot h_j]$, with $|dz|$ as their only geometric input. No
ordered concatenation, no unit direction vector, so the modulation is exchange symmetric; multiplying
an antisymmetric kernel by a symmetric modulation leaves the message antisymmetric. Hence
\begin{equation}
\eta \equiv \Big\|\sum_i \kick_i\Big\| \Big/ \sum_i\big\|\kick_i\big\| = 0
\label{eq:selfforce}
\end{equation}
identically, to floating-point round-off, for any bunch and any weights. The left-hand side is the
natural diagnostic: the fraction of per-particle kicks surviving the sum, bounded by unity, and
unlike a ratio to the typical single-particle force it does not depend on the macroparticle count.
Replace the symmetric combination with the ordered concatenation $[h_i,h_j]$---same width, a one-line
change---and antisymmetry breaks, the bunch acquires a net force on itself, and the centroid drifts
unphysically; Sec.~\ref{sec:checks} compares the two. The same argument holds inside a slice, where
the analytic field is odd in the transverse offset and the per-particle closure must therefore be
even: Eq.~(\ref{eq:edge_slice}) contains only squared offsets, the modulated kick stays odd, and each
slice exerts exactly zero force on itself---measured in Sec.~\ref{sec:checks} as a residual six
orders of magnitude below what the signed input produces. Where the interaction genuinely is ordered,
the input is ordered: Eq.~(\ref{eq:edge_wake}) presents witness and source in fixed order, which is
what lets causality be represented and the boundary condition $W_\perp(0)=0$ be preserved. For the
wake and CSR graphs a nonzero net force is the physics, since the bunch gives up momentum to the
structure and to the radiation field; the vanishing self-force of Eq.~(\ref{eq:selfforce}) belongs to
the space-charge solver alone.

\paragraph*{Silent failure modes.} Several defaults of standard graph networks are inappropriate
here, and they can fail without any visible sign in training: no divergence, an ordinary loss curve,
and a learned quantity systematically biased whenever particle number or bunch density changes.
Table~\ref{tab:silent} lists them with the choice that removes each; Supplementary
Section~\ref{sm-app:collective} gives the details.

\begin{table}[tb]
\caption{Defaults of standard graph networks that can fail silently when the network estimates a
Coulomb-like interaction in a relativistic bunch, and the choice that removes each.}
\label{tab:silent}
\begin{ruledtabular}
\footnotesize\setlength{\tabcolsep}{3pt}
\begin{tabular}{p{0.28\columnwidth}p{0.31\columnwidth}p{0.31\columnwidth}}
\raggedright Default & \raggedright Consequence & \raggedright Choice made here\tabularnewline
\colrule
\raggedright Neighbors in the laboratory frame & \raggedright different neighbors; kick off by $79\%$ & \raggedright graph on rest-frame coordinates\tabularnewline
\raggedright Ordered edge input $[h_i,h_j]$ & \raggedright net self-force, centroid drift & \raggedright symmetric input, Eq.~(\ref{eq:edge_sc})\tabularnewline
\raggedright Signed offsets in a slice closure & \raggedright slice pushes its own centroid & \raggedright squared offsets only, Eq.~(\ref{eq:edge_slice})\tabularnewline
\raggedright Symmetrized $k$-NN without de-duplication & \raggedright kick doubled, uniformly & \raggedright de-duplicated edges\tabularnewline
\raggedright Mean aggregation, graph pooling & \raggedright force diluted by $1/N$ & \raggedright sum aggregation, no pooling\tabularnewline
\raggedright All inputs normalized & \raggedright dense and relaxed bunch indistinguishable & \raggedright absolute-scale scalars kept\tabularnewline
\raggedright $u^{-2}$ as an edge feature & \raggedright $\tanh$ saturates, no gradient & \raggedright $\log u$, $1/(u^{2}+1)$, $u/(1+u)$\tabularnewline
\raggedright Fixed small softening length & \raggedright fitted emittance growth $\propto1/N$ & \raggedright adaptive softening\tabularnewline
\raggedright Undirected wake graph & \raggedright wake not representable & \raggedright causal, directed graph\tabularnewline
\end{tabular}
\end{ruledtabular}
\end{table}

\paragraph*{Bounded amplitude.} Every closure amplitude is limited and zero-initialized. The three
message closures use $\lambda=0.3$; the envelope shape factor is held tighter, to
$F\in[0.85,1.15]$. The reason for the tighter bound is specific. $F$ is a free function of
$(\sigma_x,\sigma_y,\gamma,s,I)$ and carries no information about the transverse distribution shape.
Given a wide range it will absorb quadrupole calibration error, emittance error and energy-profile
error---all functions of those same variables---instead of expressing the non-uniformity it is meant
to express. Narrowing the band from $\pm0.5$ to $\pm0.15$ leaves room for a realistic departure from
uniform density while removing the capacity to act as a general-purpose error sink.

\subsection{Identifiability by a second observable}
\label{sec:bridge_main}

Space charge and transverse wakefields both drive projected emittance growth, and the measured
growth rate gives one number. An optimizer can therefore raise one branch and lower the other at
constant loss; both then become unidentifiable, and a fitted space-charge scale cannot be read as a
statement about space charge. The two mechanisms differ in a way that can be turned into two
observables. Space charge is local---a particle feels the charge immediately around it---so growth
occurs \emph{within} a slice. A transverse dipole wake, driven by the offset of slices ahead,
displaces whole slices relative to one another; that is the banana distortion, and it can leave the
intraslice emittance untouched. The projected second-moment matrix splits exactly into the
charge-weighted mean of the intraslice matrices and the matrix of the slice centroids, giving the
emittance decomposition
\begin{equation}
\varepsilon^{2}_{\rm proj}=\varepsilon^{2}_{\rm slice}+\varepsilon^{2}_{\rm banana},
\label{eq:split}
\end{equation}
where $\varepsilon^{2}_{\rm slice}$ is the squared emittance of the bunch after each slice's own
centroid has been removed in $x$, $x'$, $y$ and $y'$, i.e. the determinant of the charge-weighted
mean of the per-slice second-moment matrices, and $\varepsilon^{2}_{\rm banana}$ is the remainder.
That remainder contains the spread of the slice centroids together with its cross term with the
intraslice moments, so Eq.~(\ref{eq:split}) holds by construction and the banana channel vanishes
when all slice centroids coincide. The slice channel is compared against the space-charge slice
component alone; the banana channel is compared against the sum of the wake banana component and the
space-charge banana component, since slices with different currents experience different defocusing
and space charge contributes there too. Identical binning and centroid removal in both decompositions
keeps any statistical bias out. The module receives gradient only through loss terms, since its kicks
are not added to the output: an emittance-growth term matching the two channels of Eq.~(\ref{eq:split}),
and a slice-energy term that is the only path to the longitudinal kicks and to the CSR kernel
(Supplementary Section~\ref{sm-app:collective}). The decomposition is a necessary condition for
identifiability, not a demonstration of it; whether an optimizer exploits it is discussed in
Sec.~\ref{sec:limits}.

\subsection{Conditioning on the machine settings}
\label{sec:generalization}

A model is first trained per working point, with the envelope residual and the collective module;
these per-working-point models, trained for 800 epochs each, form the teacher corpus. A single
student, conditioned on $w$, then reuses the envelope and centroid MLPs, adds a working-point
embedding ($\mathbb{R}^{15}\to\mathbb{R}^{32}$) upstream of both and a correlation head for
$\rho_x,\rho_y$. Each training sample is one complete trajectory, so a batch is a set of whole curves
with shape $[B,N,\cdot]$, and boundary, smoothness and sequence terms act between physically
adjacent watch points.

Supervision of the student is twofold. At the watch points it is fitted to the tracking data; between
them it is fitted to its teacher. Each teacher is rebuilt from its checkpoint, verified against the
predictions it stored at the watch points, and its pointwise path---the function the envelope
residual constrains---is evaluated on a grid between the watch points. The teacher curve is shifted
by its own error at the watch points and interpolated linearly, so the two sources of supervision
agree where they meet. The student itself evaluates neither the residual nor the collective module
(Table~\ref{tab:active}): the collective module needs a particle distribution, which does not exist
at an unseen working point, and the physics reaches the student through the teacher curves. That is
also why the student runs in $0.16$~s, and it bounds its validity to the region the corpus covers.
Input standardization of $w$ and output bounds of the student are computed from the training working
points only, so nothing of the held-out set enters the model.

The dominant failure mode of this step is a collapse of the centroid. Centroid is micron scale and
envelope is millimeter scale; scaled identically, the centroid error is four orders of magnitude
smaller. In our runs the optimizer predicted the dataset-average centroid for every working point,
erasing exactly the working-point-specific structure the model exists to provide. Rescaling the
envelope by $10^{3}$ and the centroid by $10^{4}$ removes the collapse; the centroid path is then
frozen once it has converged, at $27\%$ of the schedule, before weight averaging begins. Losses and
training schedule are in Supplementary Section~\ref{sm-app:training}.

\begin{table}[tb]
\caption{Which branch is active on which side of the distillation. The transformer is enabled in
the teacher and student runs; it is off in every single-working-point comparison, those of
Fig.~\ref{fig:collective} included. The CSR branch of the collective module is on in every run that
carries the module, except ablation (e) of Table~\ref{tab:main}, which removes it.}
\label{tab:active}
\begin{ruledtabular}
\footnotesize
\begin{tabular}{p{0.44\columnwidth}cl}
Branch & Teacher & Student\\
\colrule
Envelope and centroid MLPs & yes & yes\\
Envelope-equation residual & yes & via teacher curves\\
Adaptive region decomposition & yes & features only\\
Collective module & yes & via teacher curves\\
Transformer and fusion & yes & yes\\
Working-point embedding & --- & yes\\
Correlation head ($\rho_x,\rho_y$) & --- & yes\\
\end{tabular}
\end{ruledtabular}
\end{table}

\section{Verification without training}
\label{sec:checks}

Does the net self-force of a bunch vanish? Does a kick scale with the number of neighbors? These are
questions about the equations the network is made of, not about training, and one forward pass over
one bunch answers them. Each row of Table~\ref{tab:checks_measured} is such a pass, done twice: with
the module as built, and with the single design choice in question inverted. That control
distinguishes a structure forbidding a residual from a configuration in which the residual merely
happens to be small. The zero-initialized module reproduces the analytic kernel exactly and every
variant would agree, so the weights are randomized before each measurement; what is tested is whether
the structure protects the invariant for \emph{arbitrary} weights. Sampling noise on a tracked bunch
exceeds the effect by an order of magnitude, so each invariant is measured on a configuration
constructed to make it exact, for instance a bunch in which every particle has a mirror at $(-x,-y)$
in the same slice (Appendix~\ref{app:checks}).

These invariants belong to the network's construction, not to its training: with randomized weights
the bunch-level self-force separates by eight orders of magnitude and the slice-level one by nearly
six. Breaking the design removes the invariant rather than degrading it---the residual jumps by five
to eight orders of magnitude and its size is then set by the weights, as the scatter shows: at thirty
seeds the protected slice quantity has a relative spread of $9\%$, the broken one $234\%$ of its
mean. The recipe applies to any architecture claiming an invariant by construction: build the
configuration in which the law is exact, randomize the weights, report the broken case alongside the
intact one. These are checks, not ablations; the accuracy each choice buys is measured in
Sec.~\ref{sec:experiments}.

\begin{table}[tb]
\caption{Architectural invariants, each measured on one bunch at $N_{\rm macro}=8192$ with the
design choice inverted, over five random initializations and thirty for the slice row. $\eta$ is the
coherence fraction of Eq.~(\ref{eq:selfforce}); $\eta_{\rm slice}$ is the same quantity formed
within a slice and charge-weighted across slices.}
\label{tab:checks_measured}
\begin{ruledtabular}
\footnotesize
\begin{tabular}{p{0.33\columnwidth}p{0.19\columnwidth}p{0.32\columnwidth}}
Claim & Inverted & Measured\\
\colrule
Symmetric edge input gives zero net self-force &
ordered $[h_i,h_j]$ &
$\eta=3.3\times10^{-10}$ vs $4.1\times10^{-2}$\\[2pt]
Even slice closure gives zero slice self-force &
signed $u_x,u_y$ &
$\eta_{\rm slice}=6.4\times10^{-9}$ vs $4.7\times10^{-3}$\\[2pt]
Sum aggregation is extensive in the neighbor count &
mean aggregation &
ratio $21.4\pm2.2$ against a mean degree of $21.0$\\[2pt]
Symmetrized graph must deduplicate &
skip \texttt{unique} &
$1.90\times$ the edges, $2.00\times$ the kick in core and halo alike\\[2pt]
Rest-frame graph selects the right neighbors &
build on $(x,y,z)$ &
neighbor overlap $5.5\%$, kick deviation $79\%$\\
\end{tabular}
\end{ruledtabular}
\end{table}

\section{Experiments}
\label{sec:experiments}
\label{sec:results}

\subsection{Setup, baselines and metrics}

The quoted error is relative throughout: mean absolute error divided by the mean reference $\sigma$
over the section. The envelope passes through waists where $\sigma$ falls to $0.06$~mm, and an
absolute threshold in millimeters flatters any fit there. The baseline is a purely data-driven model,
labeled \textsc{dual} after its two prediction paths. It shares the inputs, the $R=4$ region
decomposition and the data losses, and differs in two architectural respects: one $\sigma$ trunk
shared by the two planes, and a fixed output window for $\sigma$, a sigmoid spanning $0.02$ to
$1.5$~mm, where the hybrid model derives the window from the training data, from half the smallest to
$1.5$ times the largest observed $\sigma$ (Supplementary Section~\ref{sm-app:arch}). Ablations remove
one component of the hybrid model at a time: (a)~the envelope residual, (b)~the collective module,
(c--e)~one collective branch (space charge, wakefields, CSR), and (f)~the learned region
decomposition, replaced by uniform memberships so that $K_{\rm blend}(s)$ becomes a single smooth
function of $s$ with no learned boundaries. Architecture is separated from physics by two controls:
(g)~the hybrid architecture with no physics at all, neither residual nor collective module, and
(h)~the hybrid model with one trunk shared by $\sigma_x$ and $\sigma_y$, as in \textsc{dual}. A third,
(i), gives \textsc{dual} the data-derived output window of the hybrid model. The tracking code is the
reference for all of them; what is compared is learning approaches on the same data, not a learned
model against a simulation. For the conditioned model, the student is compared with and without
distillation. All models see the same tracking data, which always contain longitudinal space charge,
short-range wakefields and CSR but no transverse space charge: removing a physics component removes it
from the model, not from the reference, so a comparison measures how well each model represents
collective effects present in the data, not how large those effects are. For the transverse
space-charge part of the module, which has no counterpart in the reference, the comparison tests
only that including it does no harm. Every per-working-point configuration is run with five seeds and the
same budget of 300 epochs, chosen for speed (Supplementary Section~\ref{sm-app:training}), so that
the configurations are compared at equal training length; entries give mean $\pm$ standard deviation
over seeds.

\begin{table*}[t]
\caption{Relative envelope error (MAE / mean $\sigma$, in \%) and centroid error. Per-working-point
models at \texttt{sxfel450}, 300 epochs, transformer off; conditioned models: error averaged over the
50 held-out working points. Mean $\pm$ standard deviation over five seeds for every entry;
$^{*}$~$p<0.05$, $^{**}$~$p<0.001$ against the full hybrid model (Welch's $t$ test); the two
conditioned models do not differ significantly.}
\label{tab:main}
\begin{ruledtabular}
\footnotesize\setlength{\tabcolsep}{4pt}
\begin{tabular}{llccc}
Setting & Model & $\sigma_x$ & $\sigma_y$ & centroid ($\mu$m)\\
\colrule
Per working point & hybrid (residual + collective module) & $4.28\pm0.62$ & $3.75\pm0.15$ & $0.19\pm0.01$ / $0.23\pm0.03$\\
 & data-driven (\textsc{dual}) & $5.38\pm0.29^{*}$ & $3.43\pm0.18^{*}$ & $0.19\pm0.00$ / $0.17\pm0.01^{*}$\\
 & (a) without envelope residual & $4.22\pm0.38$ & $3.64\pm0.10$ & $0.19\pm0.01$ / $0.25\pm0.04$\\
 & (b) without collective module & $4.53\pm0.45$ & $3.71\pm0.26$ & $0.19\pm0.01$ / $0.23\pm0.04$\\
 & (c) without space charge & $4.43\pm0.42$ & $3.82\pm0.05$ & $0.19\pm0.01$ / $0.23\pm0.04$\\
 & (d) without wakefields & $4.42\pm0.61$ & $3.70\pm0.12$ & $0.19\pm0.01$ / $0.23\pm0.05$\\
 & (e) without CSR & $4.60\pm0.69$ & $3.88\pm0.16$ & $0.19\pm0.01$ / $0.24\pm0.06$\\
 & (f) without region decomposition & $3.42\pm0.37^{*}$ & $5.41\pm0.36^{**}$ & $0.19\pm0.01$ / $0.24\pm0.04$\\
  & (g) hybrid architecture, no physics & $4.74\pm0.73$ & $3.84\pm0.34$ & $0.20\pm0.01$ / $0.24\pm0.04$\\
   & (h) hybrid, shared $\sigma$ trunk & $4.45\pm0.53$ & $3.74\pm0.20$ & $0.19\pm0.01$ / $0.25\pm0.05$\\
 & (i) \textsc{dual}, data-derived window & $3.45\pm0.37^{*}$ & $3.19\pm0.14^{**}$ & $0.19\pm0.00$ / $0.18\pm0.01^{*}$\\
\colrule
Conditioned (held out) & student, data only & $6.65\pm0.28$ & $6.71\pm0.26$ & $0.255\pm0.018$ / $0.367\pm0.040$\\
  & student, distilled & $6.75\pm0.53$ & $6.68\pm0.27$ & $0.249\pm0.012$ / $0.369\pm0.004$\\
\end{tabular}
\end{ruledtabular}
\end{table*}

\subsection{What the physics components and the architecture contribute}

\begin{figure*}[t]
\centering
\includegraphics[width=0.86\textwidth]{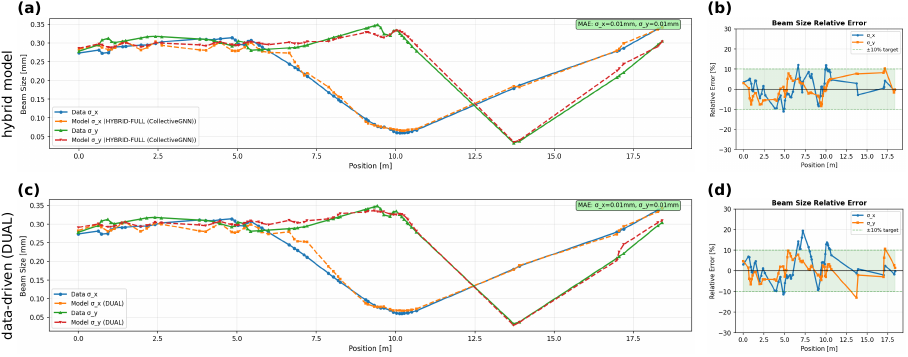}
\caption{Hybrid model against a purely data-driven one at working point \texttt{sxfel450}, 300
training epochs in both cases, seed 1; Table~\ref{tab:main} gives the statistics over five seeds.
(a),(b) the hybrid model, with the envelope-equation residual and the collective module, including
its CSR branch, active. (c),(d) the purely data-driven model, labeled \textsc{dual} after its two
prediction paths for the envelope and the centroid, with the same $R=4$ region decomposition but
trained on the data losses alone, with no physics residual and no collective module. The transformer
branch is off in both. The two models differ in architecture---above all in the output window for
$\sigma$---as well as in the physics components; Table~\ref{tab:main} separates the contributions.
Both are trained on the same \textsc{Pelegant} reference, which contains longitudinal space charge,
short-range wakefields and CSR in every case: the data-driven model lacks the collective physics as a model
component, not in its training data. Left: horizontal and vertical rms beam size, model against
\textsc{Pelegant}. Right: relative envelope error, with the $\pm10\%$ band shaded. Absolute mean
absolute error is $0.01$~mm in both runs; the relative error differs most in the horizontal plane,
where the data-driven model over-predicts $\sigma_x$ on the approach to the waist, reaching about
$+20\%$ near $s\simeq7$~m.}
\label{fig:collective}
\end{figure*}

\texttt{sxfel450} is picked at random from the training set as the main example;
Appendix~\ref{app:multiwp} repeats the key comparisons at fifteen more.
Figure~\ref{fig:collective} sets the hybrid model against the data-driven \textsc{dual} model at
\texttt{sxfel450}, with a matched budget of 300 epochs and $R=4$ regions in both. Over five seeds
(Table~\ref{tab:main}) the relative envelope error in $x$ is $4.3\pm0.6\%$ against $5.4\pm0.3\%$,
lower by a fifth for the hybrid model ($p=0.01$). The advantage sits in the drifts, which hold 51 of
the 57 watch points and the horizontal waist (Appendix~\ref{app:results}); on the approach to that
waist the data-driven run of Fig.~\ref{fig:collective} over-predicts the envelope by up to about
$20\%$, the hybrid run by at most about $12\%$. Vertically the data-driven model is the lower,
$3.4\pm0.2\%$ against $3.7\pm0.1\%$ ($p=0.02$), and it reproduces the vertical centroid somewhat
better, $0.17$ against $0.23$~$\mu$m.

\begin{figure}[tb]
\centering
\includegraphics[width=\columnwidth]{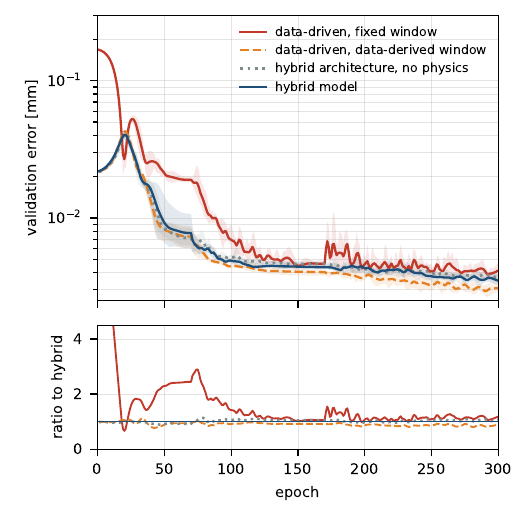}
\caption{Validation error, as defined in the text, against training epoch on a logarithmic scale,
with the ratio to the hybrid model below: the hybrid model; the same architecture without any physics
(control (g)); the data-driven \textsc{dual} model with its fixed output window for $\sigma$; and the
same data-driven model given the hybrid model's data-derived window (control (i)). Lines are means
over five seeds, shaded bands the range across seeds. Removing the physics leaves the curve within
$13\%$ of the hybrid one from epoch 10 on, and with the same output window the data-driven model
starts at the same error and converges as fast. The one curve that stands apart differs from the
others only in that window: a normalization choice, not physics.}
\label{fig:conv}
\end{figure}

Where do these differences come from? The ablations answer that. Removing the envelope residual, the
collective module, or any one collective branch changes neither accuracy (Table~\ref{tab:main};
$p>0.19$ in both planes for all five) nor convergence: validation errors at epochs 1, 50 and 300
agree with those of the full model within the seed scatter (Fig.~\ref{fig:conv}). Only the region
decomposition matters on removal. Without it the vertical error rises from $3.7\%$ to $5.4\%$
($p<0.001$), most of all at the X-band watch points ($5.4\%$ to $12.6\%$), where watch points are
sparsest and the focusing character changes; the horizontal error falls, from $4.3\%$ to $3.4\%$
($p=0.03$); and training slows markedly, the validation error at epoch 50 being three times that of
the full model. The controls settle the rest. The hybrid architecture with no physics at all, control
(g), starts and converges like the full model and reaches $4.7\pm0.7\%$ and $3.8\pm0.3\%$, not
significantly different from the full model ($p=0.32$ and $0.59$) nor, in $x$, from \textsc{dual}
($p=0.12$); the faster convergence is therefore a property of the architecture. Sharing one $\sigma$
trunk between the planes, control (h), makes no measurable difference ($4.45\pm0.53\%$ and
$3.74\pm0.20\%$, $p>0.6$), so the independent per-plane trunks are not the source. That leaves the
output window. At initialization the last layers output nearly zero and the sigmoid sits at its
midpoint, about $0.76$~mm in \textsc{dual} and about $0.27$~mm in the hybrid model, against reference
sizes of $0.03$ to $0.35$~mm at this working point; with two of the six components of the validation
vector predicted, these offsets account for the starting errors of $0.17$ and $0.022$~mm. Control
(i) isolates the window. Given the data-derived window of the hybrid model, \textsc{dual} starts at
$0.022$~mm in every seed, follows the hybrid curve throughout training (Fig.~\ref{fig:conv}), and
ends with the lowest error of all the per-working-point models, $3.45\pm0.37\%$ and $3.19\pm0.14\%$,
below the full hybrid model in both planes ($p=0.04$ and $p<0.001$) and below \textsc{dual} with its
fixed window ($p<0.001$ and $p=0.05$). The advantage of the hybrid model over the data-driven
baseline is thus accounted for by the output window, a normalization choice and not physics; with the
window equalized, the simpler data-driven model is the more accurate at this working point, by a
fifth in $x$ and a seventh in $y$. Of all the factors tested, this single normalization choice has
the largest effect on the starting error and the convergence speed, and a larger effect on the final
accuracy than any physics component: it lowers the horizontal error of the data-driven model by a
third, from $5.4\%$ to $3.5\%$, where no physics ablation moves the error of the hybrid model
measurably. Which of the hybrid model's further differences costs that accuracy---its additional loss
terms or its larger network---we have not isolated.

Repeating the key comparisons at fifteen further working points (Appendix~\ref{app:multiwp},
Tables~\ref{tab:multiwp} and \ref{tab:population}) confirms two of these findings and qualifies the
third. At none of the 16 working points do the physics components make a significant difference (the
full hybrid model against the same architecture without physics, $p\ge0.07$ at every working point,
$p=0.40$ and $0.35$ across working points), and the output window sets the starting error and the
convergence everywhere: given the data-derived window, \textsc{dual} starts within $1\%$ of the
hybrid model's error, where with its fixed window it starts $2.2$ to $7.9$ times higher, and the
window lowers the error of \textsc{dual} in both planes at all 16 working points. With equal windows,
however, the two architectures differ by plane: vertically the data-driven model is the more accurate
at eleven working points and less accurate at one (median $-27\%$, $p<0.001$ across working points),
horizontally neither is systematically better (five against three, $p=0.46$). The output window thus
accounts for the convergence advantage everywhere, while the accuracy differences between the two
architectures depend on the plane and the working point. Removing the region decomposition changes
the error significantly at most working points, by factors between $0.5$ and $2.6$, but in either
direction and with no systematic sign.

Four reasons probably explain why the physics components do not measurably change the watch-point
accuracy here. The collective module reaches the predicted envelope only through the space-charge
term of the residual, Eq.~(\ref{eq:kv}), which enters the loss as a soft constraint. The transverse
space charge that this term describes is absent from the reference, which contains longitudinal
space charge only (Sec.~\ref{sec:data}), so the transverse branch has nothing in the data to
represent. The collective effects the reference does contain are weak, the projected emittance
growing by about one percent from entrance to exit. And with 57 watch points the data terms
dominate: late in training at \texttt{sxfel450} the residual term, after its weight, is about two
orders of magnitude below the envelope data term, so its soft saturation never engages. The physics components guarantee the
structural properties of Secs.~\ref{sec:collective} and \ref{sec:checks}---conservation, bounded
departure from the analytic kernels, a second observable for identifiability---without a measurable
cost in accuracy. One might expect them to pay where data are scarce: between sparse watch points,
where the envelope residual is the only constraint, in a small corpus, and in the conditioned model,
which learns from the teachers' curves between the watch points. Neither test shows a measurable gain
in this case. Trained on 18 of the 57 watch points, with the horizontal waist of \texttt{sxfel450} in
a gap between two of them, the full hybrid model and the hybrid architecture without physics reach
the same error at the 39 watch points they never saw, $8.2\pm0.7\%$ against $8.0\pm0.4\%$ in $x$ and
$5.4\pm0.2\%$ against $5.3\pm0.4\%$ in $y$, and removing the residual or the collective module alone
does not change it measurably either (Appendix~\ref{app:thinning}). Repeated at the fifteen further
working points, the comparison with the architecture without physics comes out the same at all 16:
the error at the unseen watch points differs significantly at none of them ($p>0.13$), and the median
difference across working points is $-0.12$ percentage points in $x$ and $+0.12$ in $y$ ($p=0.14$
and $0.18$; Table~\ref{tab:sparse_multiwp}). The output window behaves differently there. It still
sets the starting error, but no longer lowers the error of \textsc{dual} at the unseen watch points
systematically, and with equal windows the hybrid architecture is the more accurate at five working
points in $x$ and six in $y$, less accurate at none and two ($p=0.16$ and $0.06$ across working
points)---vertically the reverse of the ordering with dense observation. The corpus-size test is the
subject of Sec.~\ref{sec:dataeff}.

The convergence histories of Fig.~\ref{fig:conv} tell the same story. Validation error there is the
quantity the training code monitors each epoch at the training watch points, about a third of the
envelope mean absolute error, and serves to compare the models. The hybrid model starts from
$0.022$~mm against $0.17$~mm in every seed; only at the end of the 20-epoch learning-rate warm-up,
epochs 19 to 22, does its error rise transiently above the data-driven one. Between epochs 40 and 80
the data-driven error is $1.6$ to $2.9$ times larger on average, from about epoch 120 the two are
within $20\%$, and at epoch 300 the hybrid error is the lower in every seed, $0.0035$ against
$0.0041$~mm. All five physics ablations, and the hybrid architecture without any physics, follow the
hybrid curve (validation error $0.022$, $0.008$--$0.010$ and $0.0035$--$0.0038$~mm at epochs 1, 50
and 300), so the better start and the faster convergence are properties of the architecture, and
control (i) traces them to the data-derived output window: the data-driven model given that window
starts at the same error and stays within $25\%$ of the hybrid curve from epoch 10 on, lower for
most of training; without the region decomposition the error at epoch 50 is $0.027$~mm.

\subsection{Generalization to unseen settings}

Averaged over the 50 held-out working points and five seeds, the distilled conditioned model reaches
$6.7\pm0.5\%$ in $\sigma_x$ and $6.7\pm0.3\%$ in $\sigma_y$, $0.25$ and $0.37$~$\mu$m in the
horizontal and vertical centroid, and $5.8\%$ and $8.4\%$ in the horizontal and vertical emittance
(Table~\ref{tab:main}). The student trained on the data alone is indistinguishable from it,
$6.6\pm0.3\%$ and $6.7\pm0.3\%$ ($p=0.71$ and $0.87$). Conditioning on $w$ costs accuracy: at its own
training point a per-working-point model reaches $4.3\%$ and $3.7\%$, so one model valid across the
sampled range has an envelope error about one and a half times higher in $x$ and one and four-fifths
times higher in $y$, although the comparison is not fully controlled (per-working-point figures at
one working point and 300 epochs, the student's averaged over fifty).

\begin{figure*}[t]
\centering
\includegraphics[width=0.96\textwidth]{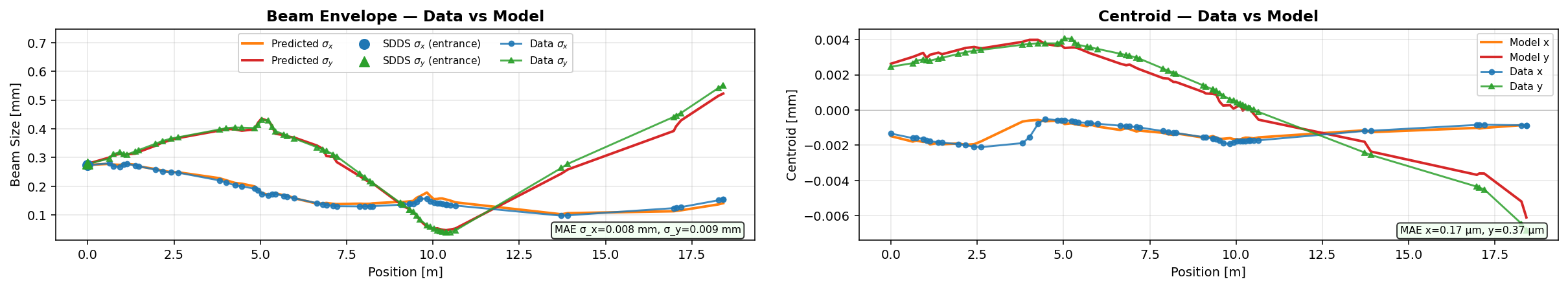}
\caption{Prediction at working point \texttt{sxfel019}, which was not used in training
(distilled student, seed 1). Left: horizontal and vertical rms beam size; solid curves are the
model, markers the \textsc{Pelegant} reference. Right: horizontal and vertical centroid. The model
receives only the 15-dimensional working-point vector, the watch-point coordinates and the fixed
entrance distribution. Mean absolute errors are $8$ and $9$~$\mu$m for $\sigma_x$ and $\sigma_y$
($4.7\%$ and $3.1\%$), and $0.17$ and $0.37$~$\mu$m for the horizontal and vertical centroid.}
\label{fig:envcent}
\end{figure*}

Figure~\ref{fig:envcent} shows the prediction at the held-out working point \texttt{sxfel019}
(distilled student, seed 1), whose errors, $4.7\%$ in $\sigma_x$ and $3.1\%$ in $\sigma_y$, lie below
the held-out average. Here, unlike \texttt{sxfel450}, the vertical plane carries the deep waist:
$\sigma_y$ falls to $0.039$~mm at $s\simeq10.4$~m and then grows to $0.55$~mm at the exit, while
$\sigma_x$ shrinks from $0.27$~mm to a broad minimum of about $0.10$~mm near $s\simeq13.7$~m, so the
model must place each waist in the correct plane from $w$ alone. It also reproduces the crossing of
the horizontal and vertical centroids between $s\simeq10.7$ and $13.7$~m---the feature that
collapses first when conditioning is inadequate---with centroid errors of $0.17$ and $0.37$~$\mu$m
against a vertical excursion of $11$~$\mu$m. Largest envelope deviations sit at the vertical waist,
where the model does not reach the full depth ($0.048$ against $0.039$~mm), and in the sparsely
observed stretch of the accelerating structures; Appendix~\ref{app:results} analyzes the emittance,
the exit parameters and the residuals at this working point.

\subsection{Data efficiency}
\label{sec:dataeff}

\begin{figure}[tb]
\centering
\includegraphics[width=\columnwidth]{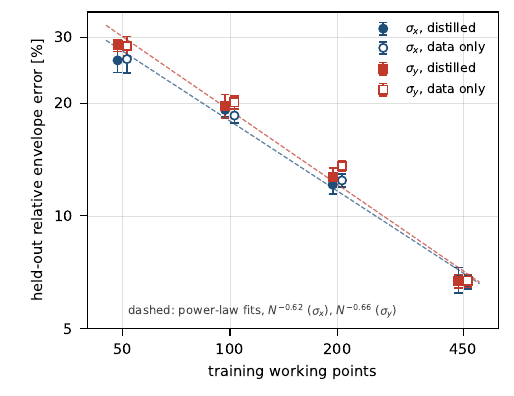}
\caption{Held-out relative envelope error of the conditioned model, averaged over the 50 held-out
working points, against the number of training working points (nested subsets). Filled markers:
distilled from the physics-constrained teachers; open markers: trained on the data alone; error
bars: standard deviation over five seeds. Dashed lines: power-law fits to the mean of both models.}
\label{fig:dataeff}
\end{figure}

Held-out error against the number of training working points appears in Fig.~\ref{fig:dataeff}, for
nested subsets of 50, 100 and 200 and the full 450, each with five seeds, with and without
distillation. The error falls as a power law, $N^{-0.62}$ in $\sigma_x$ and $N^{-0.66}$ in
$\sigma_y$, from about $26\%$ and $28\%$ at 50 working points to $6.7\%$ at 450; halving it would
take about three times as many working points. Distillation does not measurably change the error at
any corpus size here: none of the eight envelope comparisons differs significantly, the smallest $p$
being $0.06$, for $\sigma_y$ at 200 working points, where the distilled model is lower by $7\%$. A
benefit of the physics carried by the teachers would be expected to show first where data are
scarcest, at 50 and 100 working points; within the seed scatter it does not, in this case. The error
of the conditioned model on this section is governed mainly by the size of the corpus.

\subsection{Computational cost}
\label{sec:cost}

A forward pass over all 57 watch points takes $0.159$~s on a single NVIDIA  Quadro RTX 6000; the full online
path, including loading the checkpoint, $0.378$~s; the corresponding \textsc{Pelegant} run on a
16-core Intel Xeon server, $20.34$~s. The speedups compare different hardware. One-time cost of
building the corpus and the point at which it is recovered are given in Supplementary
Section~\ref{sm-app:cost}; the surrogate pays off where evaluations must be made one after another,
as in a tuning loop, and is differentiable with respect to $w$, so a setting can be optimized
directly rather than sampled. Finally, training teacher models with 500 working points takes about $100$~h on dual Quadro RTX 6000 GPUs.

\section{Discussion and limitations}
\label{sec:limits}

The design principles should carry over beyond this application. Frame, direction, aggregation and
normalization are physics choices when a graph network estimates a physical quantity, and getting
them wrong can produce errors systematic in particle number or density that the loss curve need not
show (Table~\ref{tab:silent}). A fitted parameter of a collective model cannot be interpreted
physically unless the mechanisms it distinguishes are separately identifiable, and identifiability
often has to be engineered by constructing a second observable. A conservation law imposed by
construction is a different object from one imposed by a penalty, a difference testable without any
training (Sec.~\ref{sec:checks}). The ablations also delimit what the physics components did in this
test case: on a section with weak collective effects they guaranteed these properties without a
measurable cost in accuracy or convergence, and the larger factor in training was not the physics but
a normalization choice, the data-derived output range of $\sigma$. At all 16 working points tested it
set the starting error and the convergence speed; at \texttt{sxfel450} it also accounted for the
whole accuracy advantage of the hybrid model over the data-driven baseline; across the 16 working
points the data-driven architecture given the same window was the more accurate vertically with
dense observation, with no systematic difference horizontally, while with sparse observation the
hybrid architecture tended to be the more accurate. Choices of this kind deserve the same scrutiny
as the physics; they are easily taken for granted, and here they outweighed it. A comparison between
a physics-informed model and a data-driven baseline therefore says little about the physics unless
the two share such choices; equalizing them is part of the ablation. Nor did they measurably improve
the model where data are scarce: trained on 18 of the 57 watch points, the hybrid model was no more
accurate between them than the same architecture without physics at any of the 16 working points, and
distilling the physics-constrained teachers did not measurably change the held-out error of the
conditioned model from 50 to 450 working points. On this section, then, the demonstrated value of the
physics components lies in the guarantees they provide. This is a statement about a weakly collective
regime, where the data terms can carry the envelope on their own; where collective effects
dominate---in a magnetic bunch compressor, where CSR and space charge drive strong emittance growth
and the envelope departs markedly from its linear-optics form---the residual and the collective
module may well improve accuracy, and testing that is a natural next step.

The model is also portable: distributions at the watch points, the settings, and the inputs of the
collective kernels---positions and bend angles of any dipoles, extent and cell geometry of the
accelerating structures---suffice. No complete lattice description, magnet calibration or field map
is needed. The transformer branch contributes little to the conditioned model on this densely sampled
section (the learned fusion gives it a weight of about $0.08$, between $0.06$ and $0.12$ over the ten
students of Table~\ref{tab:main}); it is retained as an option for sparse or noisy inputs.

Limitations are the following. \emph{The ground truth is the widely recognized high fidelity \textsc{elegant} particle
tracking, but still not measurement}; measured data carry noise and calibration offsets that
tracking does not, and tests on the real machine are the next step. The tracking also omits
transverse space charge, a correction of a few percent to the beam size on average
(Supplementary Section~\ref{sm-app:data}); a reference for comparison with measurements should
include it, and the transverse space-charge branch of the collective module is built for that case.
\emph{Identifiability is structural, not demonstrated}: whether an optimizer exploits
Eq.~(\ref{eq:split}) is best tested on a section with stronger collective emittance growth than this
one, where the emittance changes by about one percent from entrance to exit. No result here rests on
interpreting a fitted collective parameter; the global space-charge multiplier in particular is set
by a prior, not by the data, so it does not change the conclusion at all (Supplementary
Section~\ref{sm-sec:signal}). \emph{Physics approximations}: the slice description requires a
rest-frame aspect ratio above about 10, CSR is treated in one dimension with approximate transients,
both of them are valid in a regular room-temperature FEL linac. The short-range wakes use Bane's
fits; the S-band iris radius lies slightly below their range of validity, and in the X-band they hold
only up to $s\simeq1.6$~mm, so the X-band wake is extrapolated over about half of the bunch. However, the wakefield branch, with this extrapolated kernel, made no
measurable difference to the accuracy (Sec.~\ref{sec:experiments}); given the small size of the
X-band wake over this bunch, whose contribution comes mostly from the short distances the fits do
cover, an exact kernel is unlikely to change that. \emph{The phase spaces, including the longitudinal one, are predicted at one working point only}. At the optimal working point, a six-dimensional head reproduces the distribution along the line,
longitudinal planes included (Supplementary Fig.~\ref{sm-fig:psres}). The conditioned model does
not yet output bunch length, energy spread or chirp. On the section studied here the cost is slight.
Through a bunch compressor it is not: the longitudinal wakes of the S-band structures and the
X-band linearizer modify the energy chirp and its curvature at the compressor entrance, and hence
the compression factor, the peak current and the dispersive contribution to the beam size inside
the chicane. A surrogate of that section must predict them across working points. A systematic
study needs $10^{5}$ to $10^{6}$ macroparticles per run to keep the numerical noise low; that
multiplies the cost of the teacher corpus. The microbunching instability, which does not develop
within this section, is outside the present model. Both belong to the next phase (Supplementary
Section~\ref{sm-app:scope}). \emph{Uncertainty is not quantified}: the closure
bounds only the collective term, to within $\pm30\%$ of its analytic kernel, and places no bound on
the predicted beam. The spread over seeds indicates the training variance but is no calibrated
confidence interval, and an estimate of the kind in Ref.~\cite{garciacardona2024} would be needed for
applications such as beam-loss prediction. Uncertainty quantification is beyond the scope of this
article for now; the seed ensemble and the residuals at the held-out working points would allow it,
for instance through conformal prediction, without retraining.

These limitations define the realm in which the present conclusions hold. Nonetheless, they do not
alter the conclusions reported here at all, but determine how far those conclusions can be extended. We
therefore regard them as the next-stage research.

\section{Conclusion}
\label{sec:conclusion}

When a graph network is used as a physical solver, its topology, message symmetry, aggregation and
normalization have to be chosen by the physics. Doing so turns conservation laws into architectural
identities, verifiable with randomized weights and without any training. Built this way, collective
effects in a linac become bounded corrections to analytic kernels. They start at the analytic answer
and cannot leave its neighborhood. An exact split of the emittance growth supplies the second
observable that identifiability requires. Distillation into a model conditioned on the machine
settings gives a surrogate that predicts the beam along the line at settings it has never seen, in a
third of a second. In our test case, ablations over five seeds at 16 working points show that these
guarantees come without a measurable cost in accuracy or convergence. They also show that a
normalization choice, a data-derived output range for $\sigma$, matters more in training than any
physics component. It sets the starting error and the convergence speed at every working point
tested. The remaining accuracy differences between the hybrid and data-driven architectures depend on
the working point and on the density of observation. Within the hybrid model, the learned region
decomposition is the component that matters most for accuracy, in a direction that depends on the
working point. On this weakly collective section, the physics components did not measurably improve
accuracy, with dense or sparse watch points. Distilling the teachers into the conditioned model
likewise did not measurably change its held-out error at any corpus size. That error, about $7\%$
with 450 working points, falls roughly as the inverse $0.6$ power of the corpus size. The next steps
are calibration against measured data, a section with strong collective effects such as a magnetic
bunch compressor, where the physics components may pay off in accuracy, and the longitudinal phase
space across working points, so far predicted at the optimal working point only, together with the
microbunching instability.

\appendix

\section{Detailed results}
\label{app:results}

\subsection{Per-working-point comparison by element class}
Table~\ref{tab:byregion} breaks the error down by element class, averaged over five seeds. The
horizontal advantage of the hybrid model comes from the drifts, $4.4\%$ against $5.8\%$, which hold
51 of the 57 watch points and the horizontal waist; at the four S-band and two X-band watch points
the seed-to-seed scatter reaches one to two percentage points and the differences are within it.
Vertically the data-driven model is lower in the drifts and the S-band structures and higher at the
X-band watch points.

\begin{table}[tb]
\caption{Relative envelope error by element class for the hybrid model and the purely data-driven
one, at the working point of Fig.~\ref{fig:collective}, averaged over five seeds. $n$ is identical
between the two models, as it must be, since element class is a property of the lattice. Entries
are $x$ / $y$.}
\label{tab:byregion}
\begin{ruledtabular}
\begin{tabular}{lccc}
Class & $n$ & hybrid (\%) & \textsc{dual} (\%)\\
\colrule
drift   & 51 & 4.4 / 3.7 & 5.8 / 3.3\\
S-band  & 4  & 3.9 / 4.6 & 2.1 / 3.5\\
X-band  & 2  & 2.2 / 5.4 & 1.8 / 7.0\\
\colrule
whole section & 57 & 4.3 / 3.7 & 5.4 / 3.4\\
\end{tabular}
\end{ruledtabular}
\end{table}

\subsection{Held-out working point: emittance, exit and residuals}

\begin{figure*}[t]
\centering
\includegraphics[width=0.80\textwidth]{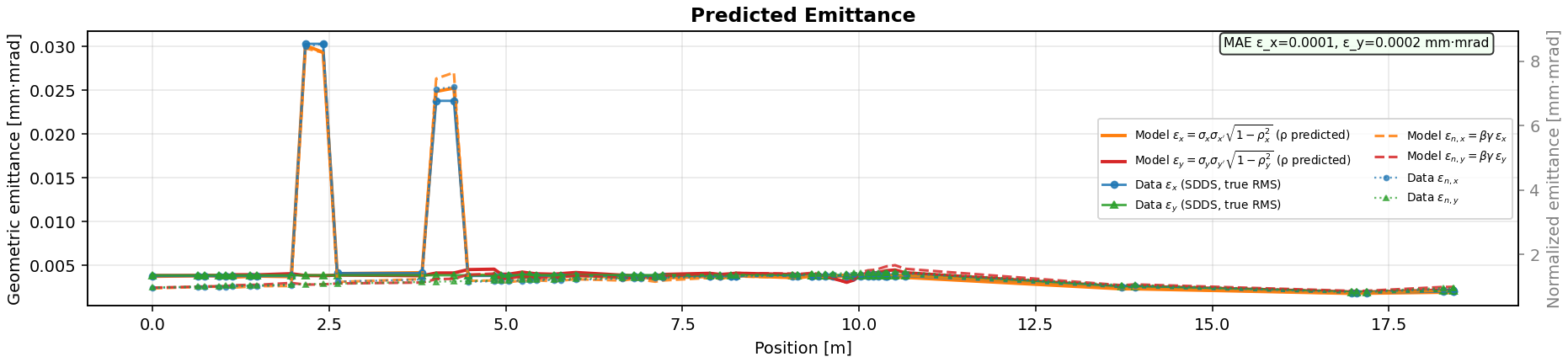}
\caption{Geometric (left axis) and normalized (right axis) emittance at the held-out working point
\texttt{sxfel019} (distilled student, seed 1). Model emittances are assembled from the predicted
$\sigma$, $\sigma'$ and $\rho$ through Eq.~(\ref{eq:emit}), not predicted directly. The two
excursions near $s\simeq2.3$ and $4.2$~m bracket the laser heater; they rise an order of magnitude
above the baseline and are reproduced rather than smoothed away.}
\label{fig:emit}
\end{figure*}

The model predicts $\rho$ and assembles $\varepsilon$ through Eq.~(\ref{eq:emit}); the emittance in
Fig.~\ref{fig:emit} therefore tests the internal consistency of the four-plane output rather than
being an independent fit. Dispersion inside the laser-heater chicane produces the two
order-of-magnitude excursions, a good test of the smoothness penalties, since a term that suppressed
them would be indistinguishable from one that suppressed genuine ripple. Over the section the mean
absolute errors are $1.5\times10^{-4}$ and $1.8\times10^{-4}$~mm\,mrad, or $2.9\%$ and $5.0\%$
relative. At the exit the model gives $\enx=0.855$ and $\eny=0.988$~mm\,mrad, $6.7\%$ below the
tracking reference of $0.916$~mm\,mrad horizontally and $3.0\%$ above the reference of
$0.959$~mm\,mrad vertically. The horizontal deviation has the same sign as the envelope deviation
there, $\sigma_x$ being $8.2\%$ low at the exit; vertically $\sigma_y$ is $5.2\%$ low while the
emittance is $3.0\%$ high, so the predicted divergence and correlation more than offset the envelope
error. Because $\varepsilon$ is assembled from three separately predicted quantities, agreement in
$\varepsilon$ at one position does not by itself imply agreement in each factor. Table~\ref{tab:exit}
collects the exit comparison.

\begin{table}[tb]
\caption{Exit parameters at the held-out working point \texttt{sxfel019}
($\beta\gamma:254.1\rightarrow463.8$; entrance $\sigma_x=0.274$, $\sigma_y=0.278$~mm; entrance
normalized emittance $0.953$ and $0.975$~mm\,mrad, from the covariance of the entrance
distribution). Model and reference values at the last watch point are taken from the residual output
of the prediction script; the reference emittance is normalized with the same $\beta\gamma$ as the
model's.}
\label{tab:exit}
\begin{ruledtabular}
\footnotesize\setlength{\tabcolsep}{3pt}
\begin{tabular}{lccc}
Quantity & Model & Ref. & Deviation\\
\colrule
$\sigma_x$ (mm) & 0.142 & 0.154 & $-8.2\%$\\
$\sigma_y$ (mm) & 0.523 & 0.551 & $-5.2\%$\\
growth $\sigma_x$ / $\sigma_y$ & $0.52\times$ / $1.88\times$ & $0.56\times$ / $1.98\times$ & \\
$\enx$ (mm\,mrad) & 0.855 & 0.916 & $-6.7\%$\\
$\eny$ (mm\,mrad) & 0.988 & 0.959 & $+3.0\%$\\
\colrule
\multicolumn{4}{l}{Whole-section error}\\
$\sigma_x$ / $\sigma_y$ & \multicolumn{3}{l}{$8.1$ / $8.6$~$\mu$m; $4.7\%$ / $3.1\%$}\\
$\langle x\rangle$ / $\langle y\rangle$ & \multicolumn{3}{l}{$0.17$ / $0.37$~$\mu$m}\\
$\varepsilon_x$ / $\varepsilon_y$ & \multicolumn{3}{l}{$1.5$ / $1.8\times10^{-4}$~mm\,mrad;
$2.9\%$ / $5.0\%$}\\
\end{tabular}
\end{ruledtabular}
\end{table}

\begin{figure*}[t]
\centering
\includegraphics[width=0.80\textwidth]{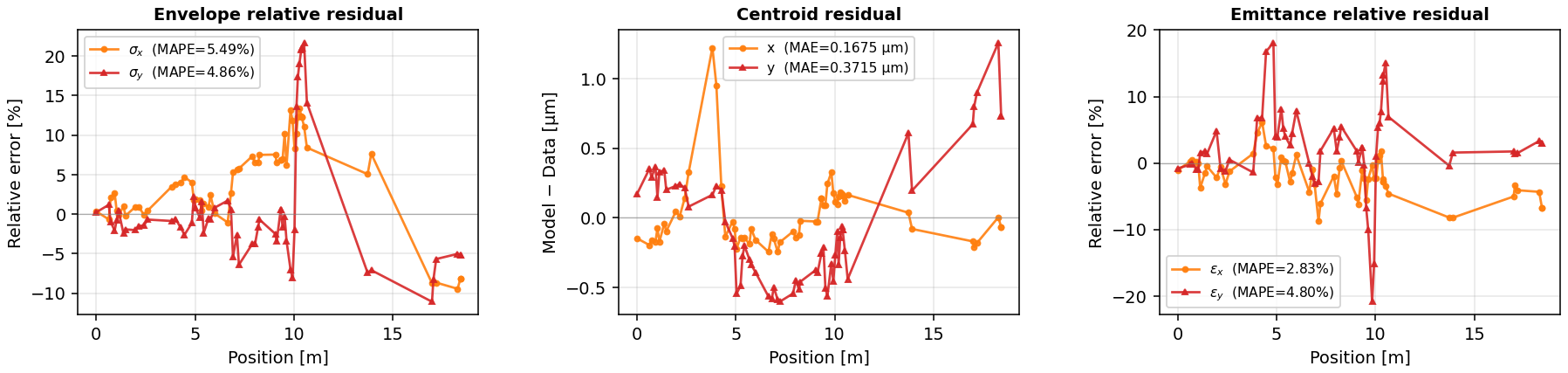}
\caption{Residuals at the held-out working point \texttt{sxfel019}. Envelope and emittance are shown
as relative errors, the centroid in absolute units because it has no meaningful denominator. Up to
$s\simeq9.4$~m the envelope error stays within $-6.4\%$ and $+7.5\%$; the excursions to $+13\%$ in
$\sigma_x$ and $+22\%$ in $\sigma_y$ between $s\simeq9.5$ and $10.7$~m lie at and around the vertical
waist, where $\sigma_y$ falls to $0.039$~mm. Beyond $s\simeq10.7$~m both planes drift low, to about
$-5$ to $-11\%$, in the stretch occupied by the accelerating structures, where watch points exist
only in the gaps between structures and at the exit. The legends quote the mean of the pointwise
relative error, whereas the relative errors in the text are the mean absolute error divided by the
mean reference value over the section, the definition used throughout.}
\label{fig:resid}
\end{figure*}

The residuals in Fig.~\ref{fig:resid} localize the error. From the entrance to $s\simeq9.4$~m the
envelope relative error stays between $-6.4\%$ and $+7.5\%$. Where the denominator is small,
excursions grow: at the vertical waist, $s\simeq10.1$--$10.7$~m, $\sigma_y$ is up to $21.7\%$ high,
because the model does not reach the full depth of the waist ($0.048$ against $0.039$~mm), and in the
same stretch $\sigma_x$ is up to $13.4\%$ high, where the model overshoots a small local bump in the
reference near $s\simeq9.9$~m. Beyond $s\simeq10.7$~m both planes end up offset rather than
scattered: $\sigma_x$ is $5$ to $8\%$ high next to its broad minimum near $s\simeq13.8$~m and $8$ to
$9\%$ low over the last meter and a half, and $\sigma_y$ is $5$ to $11\%$ low throughout. That part
has a physical cause rather than a small denominator---the watch points there sit in three close
groups up to $3.1$~m apart, so the model must span gaps comparable to the betatron wavelength, and a
model bridging an unconstrained interval is expected to end up offset; the offset carries over to the
horizontal exit emittance. The centroid residual stays within $1.3$~$\mu$m everywhere, with its
largest values just downstream of the laser heater and at the exit. Emittance relative residual is
the noisiest of the three, ranging from $-8.8\%$ to $+6.1\%$ in $\varepsilon_x$ and from $-21\%$ to
$+18\%$ in $\varepsilon_y$, the vertical extremes at the end of the laser heater and at the vertical
waist; assembled from three separately predicted quantities, $\varepsilon$ accumulates their errors
rather than averaging them.

\section{Further working points}
\label{app:multiwp}

The per-working-point comparisons of Sec.~\ref{sec:experiments} are made at \texttt{sxfel450}. To
test how far they carry over, the key configurations---the full hybrid model, the same architecture
without physics (control (g)), \textsc{dual} with its fixed and with the data-derived output window
(control (i)), and the hybrid model without the region decomposition (ablation (f))---were repeated
at fifteen further working points with the settings of Table~\ref{tab:main}, five seeds each.
Table~\ref{tab:multiwp} gives the result per working point and Table~\ref{tab:population} across
working points, each counting once. Removing the physics components makes no significant difference
at any of the 16 working points. The output window sets the starting error everywhere: with the
data-derived window \textsc{dual} starts within $1\%$ of the hybrid model and with its fixed window
$2.2$ to $7.9$ times higher, and the window lowers the error of \textsc{dual} in both planes at every
working point. With equal windows the data-driven architecture is the more accurate vertically at
most working points and neither architecture is systematically better horizontally; the region
decomposition changes the error strongly, but with a sign that depends on the plane and the working
point.

\begin{table*}[tb]
\caption{Key per-working-point comparisons at 16 working points: relative envelope error (MAE / mean
$\sigma$, in \%), $\sigma_x$ / $\sigma_y$, mean over five seeds, 300 epochs. $^{*}$~$p<0.05$,
$^{**}$~$p<0.001$ against the full hybrid model at the same working point (Welch's $t$ test).}
\label{tab:multiwp}
\begin{ruledtabular}
\footnotesize\setlength{\tabcolsep}{4pt}
\begin{tabular}{lccccc}
Working point & hybrid & hybrid, no physics & \textsc{dual}, fixed window & \textsc{dual}, data-derived window & without regions\\
\colrule
\texttt{sxfel001} & $4.39$ / $2.03$ & $3.53$ / $2.29$ & $3.12^{*}$ / $2.38$ & $2.33^{**}$ / $1.79$ & $2.68^{*}$ / $2.99^{**}$\\
\texttt{sxfel016} & $2.02$ / $4.47$ & $2.22$ / $4.17$ & $1.98$ / $4.05$ & $1.62$ / $2.78^{**}$ & $2.28$ / $4.33$\\
\texttt{sxfel019} & $2.96$ / $9.48$ & $2.84$ / $9.44$ & $5.66^{**}$ / $7.38$ & $4.52^{*}$ / $6.66^{*}$ & $3.74^{*}$ / $4.84^{**}$\\
\texttt{sxfel048} & $3.36$ / $4.72$ & $3.46$ / $4.66$ & $6.18^{**}$ / $4.52$ & $5.44^{**}$ / $4.34$ & $4.20^{*}$ / $12.07^{*}$\\
\texttt{sxfel075} & $3.23$ / $1.73$ & $3.15$ / $1.87$ & $3.49$ / $3.28^{**}$ & $2.73$ / $2.31^{*}$ & $3.39$ / $3.73^{**}$\\
\texttt{sxfel121} & $3.43$ / $3.68$ & $3.46$ / $3.92$ & $4.86^{*}$ / $4.53^{*}$ & $2.58^{*}$ / $3.25$ & $2.63^{*}$ / $3.87$\\
\texttt{sxfel179} & $2.41$ / $5.86$ & $2.62$ / $5.67$ & $3.80^{**}$ / $5.34^{*}$ & $2.55$ / $5.23^{*}$ & $3.92^{*}$ / $5.37$\\
\texttt{sxfel223} & $2.74$ / $6.66$ & $2.82$ / $7.09$ & $4.85^{**}$ / $7.62$ & $2.46$ / $3.43^{**}$ & $3.53^{*}$ / $3.95^{*}$\\
\texttt{sxfel245} & $2.17$ / $4.52$ & $2.17$ / $4.04$ & $2.77^{*}$ / $3.73$ & $2.05$ / $3.12^{*}$ & $2.75^{*}$ / $4.19$\\
\texttt{sxfel281} & $2.26$ / $5.14$ & $2.22$ / $4.95$ & $2.58^{*}$ / $3.90^{*}$ & $2.31$ / $3.43^{*}$ & $4.54^{**}$ / $5.12$\\
\texttt{sxfel317} & $2.34$ / $3.58$ & $2.77$ / $3.42$ & $5.42^{**}$ / $3.62$ & $2.53$ / $2.52^{*}$ & $4.08^{**}$ / $3.94$\\
\texttt{sxfel338} & $3.05$ / $4.13$ & $3.03$ / $4.27$ & $3.10$ / $5.67^{**}$ & $2.27^{**}$ / $3.15^{*}$ & $3.26$ / $3.18^{*}$\\
\texttt{sxfel397} & $3.29$ / $4.29$ & $3.55$ / $5.46$ & $5.05^{*}$ / $6.79^{**}$ & $3.74$ / $3.51$ & $3.44$ / $5.35$\\
\texttt{sxfel416} & $2.90$ / $5.26$ & $2.62$ / $5.58$ & $2.47$ / $3.96^{*}$ & $2.32^{*}$ / $3.60^{*}$ & $3.29$ / $4.55$\\
\texttt{sxfel450} & $4.28$ / $3.75$ & $4.74$ / $3.84$ & $5.38^{*}$ / $3.43^{*}$ & $3.45^{*}$ / $3.19^{**}$ & $3.42^{*}$ / $5.41^{**}$\\
\texttt{sxfel488} & $1.82$ / $12.30$ & $1.94$ / $12.88$ & $2.44^{*}$ / $4.37^{**}$ & $2.31^{**}$ / $3.92^{**}$ & $2.42^{**}$ / $5.85^{**}$\\
\end{tabular}
\end{ruledtabular}
\end{table*}

\begin{table*}[tb]
\caption{Population view of Table~\ref{tab:multiwp}: each working point counts once, with its mean
over seeds. For each plane, the median relative difference to the full hybrid model, the number of
working points where the model is significantly lower / higher (Welch's $t$ test over seeds,
$p<0.05$), and Wilcoxon's signed-rank test across the 16 working points; last column, the median
ratio of the starting validation error to that of the hybrid model.}
\label{tab:population}
\begin{ruledtabular}
\footnotesize\setlength{\tabcolsep}{4pt}
\begin{tabular}{lccccccc}
 & \multicolumn{3}{c}{$\sigma_x$} & \multicolumn{3}{c}{$\sigma_y$} & \\
Model & median & lower / higher & $p$ & median & lower / higher & $p$ & start error ratio\\
\colrule
hybrid architecture, no physics & $+2\%$ & 0 / 0 & $0.404$ & $+3\%$ & 0 / 0 & $0.348$ & $1.01$\\
\textsc{dual}, fixed window & $+31\%$ & 1 / 11 & $0.004$ & $-6\%$ & 5 / 4 & $0.782$ & $4.37$\\
\textsc{dual}, data-derived window & $-8\%$ & 5 / 3 & $0.464$ & $-27\%$ & 11 / 1 & $3.1\times10^{-4}$ & $1.00$\\
hybrid without region decomposition & $+19\%$ & 3 / 8 & $0.105$ & $-2\%$ & 4 / 4 & $0.900$ & $1.00$\\
\end{tabular}
\end{ruledtabular}
\end{table*}

\section{Sparse observation}
\label{app:thinning}

The 57 watch points are denser than the beam diagnostics of a real installation, and the envelope
residual constrains the model precisely between them. The per-working-point comparison of
Sec.~\ref{sec:experiments} is therefore repeated with a sparse layout. Of the 57 watch points, 18 are
kept for training by a minimum-spacing rule with both ends always kept: stations are removed only
where they are closer together than the target spacing, so the dense clusters thin first and the
sparse downstream stretch is untouched. Retained positions are $0$, $0.63$, $1.38$, $1.96$, $2.61$,
$3.81$, $4.46$, $5.23$, $5.99$, $6.64$, $7.88$, $9.05$, $9.83$, $10.51$, $13.72$, $16.98$, $18.27$
and $18.41$~m; the horizontal waist of \texttt{sxfel450}, near $s\simeq10.2$~m, where the beam is
smallest, lies in the gap between $9.83$ and $10.51$~m, which makes this a favourable case for the
envelope residual, the only constraint acting between watch points. Everything derived from the beam data---training targets, normalization, bunch
length and current profile, initial region boundaries---uses the 18 retained watch points only; the
energy profile, which a machine knows from its rf settings, is taken from the full series. Every
model is then evaluated at all 57 watch points, the 18 it was trained on and the 39 it never saw.
Settings otherwise follow Table~\ref{tab:main}: \texttt{sxfel450}, 300 epochs, $R=4$, transformer
off, five seeds per configuration.

\begin{table*}[tb]
\caption{Relative envelope error (MAE / mean $\sigma$, in \%) with 18 of the 57 watch points used in
training, at the retained (trained) and the remaining (unseen) watch points; \texttt{sxfel450}, mean
$\pm$ standard deviation over five seeds. $^{*}$~$p<0.05$ against the full hybrid model (Welch's $t$
test); no other difference is significant.}
\label{tab:sparse}
\begin{ruledtabular}
\begin{tabular}{lcccc}
 & \multicolumn{2}{c}{trained (18)} & \multicolumn{2}{c}{unseen (39)}\\
Model & $\sigma_x$ & $\sigma_y$ & $\sigma_x$ & $\sigma_y$\\
\colrule
hybrid (residual + collective module) & $2.7\pm0.4$ & $2.7\pm0.2$ & $8.2\pm0.7$ & $5.4\pm0.2$\\
without envelope residual             & $2.8\pm0.4$ & $2.7\pm0.1$ & $8.1\pm0.4$ & $5.2\pm0.2$\\
without collective module             & $2.9\pm0.4$ & $2.6\pm0.5$ & $8.2\pm0.4$ & $5.4\pm0.4$\\
hybrid architecture, no physics       & $2.6\pm0.3$ & $2.7\pm0.6$ & $8.0\pm0.4$ & $5.3\pm0.4$\\
data-driven (\textsc{dual})           & $3.1\pm0.3$ & $2.9\pm0.3$ & $7.6\pm0.6$ & $5.8\pm0.2^{*}$\\
\end{tabular}
\end{ruledtabular}
\end{table*}

Table~\ref{tab:sparse} gives the result. At the unseen watch points the error is about three times
that at the trained ones in $x$ and twice in $y$---and it is there that a constraint acting between
watch points would show. None does so measurably in this case: removing the envelope residual, the
collective module, or both does not measurably change the error at the unseen watch points
($p\ge0.40$ in every case), the horizontal waist inside an unobserved gap included. Only one
difference reaches significance: the data-driven model's vertical error at the unseen watch points,
higher by about a tenth, whereas with all 57 watch points it was the lower; its horizontal error, a
fifth higher with dense observation, is not distinguishable from the hybrid model's here.

The comparison of the full hybrid model with the hybrid architecture without physics and with
\textsc{dual} in its fixed output window was repeated, with the same thinning rule, at the fifteen
further working points of Appendix~\ref{app:multiwp}, five seeds each and otherwise with the
settings above (Table~\ref{tab:sparse_multiwp}). The error at the unseen watch points depends
strongly on the working point, from $3.5$ to $12.6\%$ in $x$ and from $4.8$ to $23.4\%$ in $y$ for
the hybrid model. Removing the physics components makes no significant difference at any of the 16
working points ($p>0.13$); across working points the median difference to the full hybrid model is
$-0.12$ percentage points in $x$ and $+0.12$ in $y$ (Wilcoxon's signed-rank test, $p=0.14$ and
$0.18$), and the starting validation error stays within $1.5\%$ of the hybrid model's at every
working point. \textsc{dual} with its fixed window starts $2.2$ to $8.5$ times higher (median
$4.7$), as with dense observation; at the unseen watch points it is significantly less accurate than
the hybrid model at four working points in $x$ and six in $y$, and more accurate at none in $x$ and
two in $y$, with no systematic difference across working points (median differences $+0.17$ and
$+0.44$ percentage points, $p=0.32$ and $0.18$). Control (i), \textsc{dual} given the data-derived
window, starts within $1\%$ of the hybrid model at every working point, so the window sets the
starting error with sparse observation as well. At the unseen watch points, however, it no longer
lowers the error of \textsc{dual} systematically: against the fixed window, the error is
significantly lower at three working points and higher at two in $x$, lower at three and higher at
four in $y$ ($^{\dagger}$ in Table~\ref{tab:sparse_multiwp}; $p=0.46$ and $0.56$ across working
points). With equal windows the hybrid architecture tends to be the more accurate between the watch
points: \textsc{dual} is significantly less accurate than the hybrid model at five working points in
$x$ and six in $y$, and more accurate at none and two (median differences $+0.24$ and $+0.75$
percentage points, $p=0.16$ and $0.06$ across working points). Vertically this reverses the ordering
found with dense observation, where \textsc{dual} with the same window was the more accurate at
eleven working points. Since removing the physics components changes nothing here, the difference
lies in the remaining architectural differences between the two models, which we have not isolated.

\begin{table*}[tb]
\caption{Sparse observation at 16 working points: relative envelope error (MAE / mean $\sigma$, in
\%) at the 39 unseen watch points, with 18 of the 57 used in training; $\sigma_x$ / $\sigma_y$, mean
over five seeds, 300 epochs. $^{*}$~$p<0.05$, $^{**}$~$p<0.001$ against the full hybrid model at the
same working point (Welch's $t$ test); $^{\dagger}$~$p<0.05$, $^{\dagger\dagger}$~$p<0.001$ for
\textsc{dual} with the data-derived window against \textsc{dual} with its fixed window.}
\label{tab:sparse_multiwp}
\begin{ruledtabular}
\footnotesize\setlength{\tabcolsep}{4pt}
\begin{tabular}{lcccc}
Working point & hybrid & hybrid, no physics & \textsc{dual}, fixed window & \textsc{dual}, data-derived window\\
\colrule
\texttt{sxfel001} & $12.64$ / $4.88$ & $12.95$ / $5.25$ & $11.97$ / $5.72^{*}$ & $12.47$ / $6.21^{*}$\\
\texttt{sxfel016} & $5.31$ / $23.41$ & $5.21$ / $23.87$ & $5.08$ / $24.74^{*}$ & $5.41$ / $20.32^{**\dagger\dagger}$\\
\texttt{sxfel019} & $7.31$ / $13.90$ & $6.50$ / $13.57$ & $8.01$ / $13.87$ & $9.78^{*}$ / $17.41^{*\dagger}$\\
\texttt{sxfel048} & $5.03$ / $14.92$ & $4.96$ / $15.01$ & $11.07^{**}$ / $15.99^{*}$ & $8.19^{*\dagger}$ / $16.04^{*}$\\
\texttt{sxfel075} & $3.53$ / $4.84$ & $3.39$ / $5.25$ & $3.94$ / $5.46$ & $3.86^{*}$ / $5.07$\\
\texttt{sxfel121} & $6.49$ / $7.91$ & $6.66$ / $7.99$ & $6.42$ / $7.22^{*}$ & $6.96^{*}$ / $8.30^{*\dagger\dagger}$\\
\texttt{sxfel179} & $10.35$ / $19.70$ & $10.28$ / $20.16$ & $13.34^{*}$ / $19.24$ & $10.58^{\dagger}$ / $24.35^{*\dagger\dagger}$\\
\texttt{sxfel223} & $7.76$ / $10.66$ & $7.74$ / $11.02$ & $9.28^{*}$ / $10.35$ & $8.21$ / $10.45$\\
\texttt{sxfel245} & $8.61$ / $10.80$ & $7.70$ / $10.93$ & $7.42$ / $13.42^{*}$ & $6.79$ / $10.96^{\dagger}$\\
\texttt{sxfel281} & $11.03$ / $17.13$ & $10.82$ / $16.05$ & $11.47$ / $16.94$ & $10.88$ / $19.36$\\
\texttt{sxfel317} & $4.90$ / $6.44$ & $5.03$ / $6.68$ & $6.06$ / $7.70$ & $3.99^{\dagger}$ / $6.16^{\dagger}$\\
\texttt{sxfel338} & $5.09$ / $8.03$ & $4.94$ / $8.47$ & $4.56$ / $8.47$ & $5.40^{\dagger}$ / $9.55^{*\dagger}$\\
\texttt{sxfel397} & $7.39$ / $19.30$ & $7.66$ / $18.41$ & $9.38^{*}$ / $19.21$ & $8.70^{*}$ / $20.83$\\
\texttt{sxfel416} & $9.73$ / $11.78$ & $9.21$ / $11.82$ & $9.28$ / $18.54^{**}$ & $9.69$ / $14.97$\\
\texttt{sxfel450} & $8.21$ / $5.36$ & $8.04$ / $5.29$ & $7.63$ / $5.81^{*}$ & $7.97$ / $5.58$\\
\texttt{sxfel488} & $5.81$ / $22.42$ & $5.66$ / $22.51$ & $5.41$ / $18.95^{*}$ & $6.07^{\dagger}$ / $20.82^{*}$\\
\end{tabular}
\end{ruledtabular}
\end{table*}

With dense or sparse observation alike, the physics components did not measurably change the
accuracy on this weakly collective section. The result is specific to that regime; a stronger weight
on the residual, a hard constraint, and above all a section with strong collective effects, such as a
magnetic bunch compressor, are the natural next tests.

\section{Construction of the architectural checks}
\label{app:checks}

The invariants cannot be read off a tracked bunch directly. The analytic field of a slice is odd in
the offset from the slice centroid, and an even modulation keeps the kick odd, but a finite sum over
a slice's particles vanishes only if the sampled offsets are symmetric too; on a tracked bunch the
invariant sits under a sampling noise an order of magnitude larger. Each particle is therefore paired
with a mirror at $(-x,-y)$ sharing its arrival time, hence its slice, so that the slice sum vanishes
to round-off if and only if the kick is odd; the first row rests on the same idea. Inverting the
slice closure has to be done inside the module, since the squared offsets destroy the sign before the
network sees them; the two parities take inputs of the same width, so the same weights serve both.

The frame comparison is the most direct: the laboratory-frame and rest-frame neighbor sets overlap by
only $5.5\%$, and using the laboratory frame changes the kick by $79\%$, with no scatter across
initializations since the overlap is pure geometry.

One claim of Sec.~\ref{sec:symmetry} is deliberately absent. Whether the absolute-scale scalars let
the closure distinguish a dense bunch from a relaxed one is not an empirical question at this level,
since a function of exclusively normalized inputs is scale invariant by algebra; the measurement we
did construct, the response to a two-fold longitudinal compression with the scalars enabled and
disabled, returned $0.013\pm0.015$, consistent with zero at five seeds. Whether a \emph{trained}
closure uses them requires a trained model and a different experiment.

\begin{acknowledgments}
This work is supported by the National Natural Science Foundation of China (Grants No.~12625511 and
No.~12435011), the CAS Project for Young Scientists in Basic Research (Grant No.~YSBR-115), the
Shanghai Municipal Science and Technology Major Project, and the Shanghai Action Plan for Science,
Technology and Innovation (Grant No.~24JD1402800).

\end{acknowledgments}


\begin{thebibliography}{99}

\bibitem{raissi2019} M. Raissi, P. Perdikaris, and G. E. Karniadakis,
\emph{Physics-informed neural networks: A deep learning framework for solving forward and inverse
problems involving nonlinear partial differential equations},
J. Comput. Phys. \textbf{378}, 686 (2019).

\bibitem{greydanus2019} S. Greydanus, M. Dzamba, and J. Yosinski,
Hamiltonian neural networks, in \emph{Advances in Neural Information Processing Systems 32} (2019).

\bibitem{cranmer2020} M. Cranmer, S. Greydanus, S. Hoyer, P. Battaglia, D. Spergel, and S. Ho,
Lagrangian neural networks, arXiv:2003.04630 (2020).

\bibitem{satorras2021} V. G. Satorras, E. Hoogeboom, and M. Welling,
E($n$) equivariant graph neural networks, in \emph{Proceedings of the 38th International Conference
on Machine Learning}, Proceedings of Machine Learning Research Vol.~139 (PMLR, 2021).

\bibitem{sanchezgonzalez2020} A. Sanchez-Gonzalez \emph{et al.},
\emph{Learning to simulate complex physics with graph networks}, in \emph{Proceedings of the 37th
International Conference on Machine Learning}, Proceedings of Machine Learning Research Vol.~119
(PMLR, 2020), p.~8459.

\bibitem{sharma2026} V. Sharma and O. Fink,
\emph{A physics-informed graph neural network conserving linear and angular momentum for dynamical
systems}, Nat. Commun. \textbf{17}, 1045 (2026).

\bibitem{rackauckas2020} C. Rackauckas, Y. Ma, J. Martensen, C. Warner, K. Zubov, R. Supekar,
D. Skinner, A. Ramadhan, and A. Edelman, Universal differential equations for scientific machine
learning, arXiv:2001.04385 (2020).

\bibitem{li2021fno} Z. Li, N. Kovachki, K. Azizzadenesheli, B. Liu, K. Bhattacharya, A. Stuart,
and A. Anandkumar, Fourier neural operator for parametric partial differential equations, in
\emph{International Conference on Learning Representations} (2021).

\bibitem{borland2000} M. Borland,
\emph{elegant: A Flexible SDDS-Compliant Code for Accelerator Simulation},
Advanced Photon Source Report No.\ LS-287 (2000).

\bibitem{wang2006} Y. Wang and M. Borland,
\emph{Pelegant: A parallel accelerator simulation code for electron generation and tracking},
AIP Conf. Proc. \textbf{877}, 241 (2006).

\bibitem{floettmann2017astra} K. Fl\"ottmann,
\emph{ASTRA: A Space Charge Tracking Algorithm}, User Manual Version~3.2, DESY, Hamburg (2017).

\bibitem{qiang2000impact} J. Qiang, R. D. Ryne, S. Habib, and V. Decyk,
\emph{An object-oriented parallel particle-in-cell code for beam dynamics simulation in linear
accelerators}, J. Comput. Phys. \textbf{163}, 434 (2000).

\bibitem{qiang2006impactt} J. Qiang, S. Lidia, R. D. Ryne, and C. Limborg-Deprey,
\emph{Three-dimensional quasistatic model for high brightness beam dynamics simulation},
Phys. Rev. ST Accel. Beams \textbf{9}, 044204 (2006).

\bibitem{edelen2024} A. Edelen and X. Huang,
\emph{Machine learning for design and control of particle accelerators: A look backward and forward},
Annu. Rev. Nucl. Part. Sci. \textbf{74}, 557 (2024).

\bibitem{ivanov2020} A. Ivanov and I. Agapov,
\emph{Physics-based deep neural networks for beam dynamics in charged particle accelerators},
Phys. Rev. Accel. Beams \textbf{23}, 074601 (2020).

\bibitem{garciacardona2024} C. Garcia-Cardona and A. Scheinker,
\emph{Machine learning surrogate for charged particle beam dynamics with space charge based on a
recurrent neural network with aleatoric uncertainty},
Phys. Rev. Accel. Beams \textbf{27}, 024601 (2024).

\bibitem{huang2023} C.-K. Huang \emph{et al.},
\emph{Symplectic neural surrogate models for beam dynamics},
in \emph{Proceedings of IPAC'23}, Venice (JACoW, Geneva, 2023), WEPA078.

\bibitem{sun2023} K. Sun, X. Chen, X. Zhao, X. Qi, Z. Wang, and Y. He,
\emph{Surrogate model of particle accelerators using encoder-decoder neural networks with physical
regularization}, Int. J. Mod. Phys. A \textbf{38}, 2350145 (2023).

\bibitem{edelen2020} A. Edelen, N. Neveu, M. Frey, Y. Huber, C. Mayes, and A. Adelmann,
\emph{Machine learning for orders of magnitude speedup in multiobjective optimization of particle
accelerator systems},
Phys. Rev. Accel. Beams \textbf{23}, 044601 (2020).

\bibitem{gupta2021} L. Gupta \emph{et al.},
\emph{Improving surrogate model accuracy for the LCLS-II injector frontend using convolutional
neural networks and transfer learning}, Mach. Learn.: Sci. Technol. \textbf{2}, 045025 (2021).

\bibitem{wan2025} J. Wan, J. Qiang, and Y. Hao,
\emph{Symplectic machine learning model for fast simulation of space-charge effects},
Phys. Rev. Accel. Beams \textbf{28}, 074602 (2025).

\bibitem{leon2025} C. Leon, P. M. Anisimov, N. Yampolsky, and A. Scheinker,
\emph{Using convolutional neural networks to accelerate three-dimensional coherent synchrotron
radiation computations}, Phys. Rev. Accel. Beams \textbf{28}, 110701 (2025).

\bibitem{zhao2017} Z. Zhao, D. Wang, Q. Gu, L. Yin, M. Gu, Y. Leng, and B. Liu,
\emph{Status of the SXFEL facility}, Appl. Sci. \textbf{7}, 607 (2017).

\bibitem{borland2008} M. Borland,
\emph{Modeling of the microbunching instability},
Phys. Rev. ST Accel. Beams \textbf{11}, 030701 (2008).

\bibitem{huangsxfel2011} D. Huang and Q. Gu,
\emph{Study of the microbunching instability in the linac of the future Shanghai soft X-ray FEL
facility (SXFEL)}, in \emph{Proceedings of FEL2011}, Shanghai, China (JACoW, 2011), THPB09, p.~579.

\bibitem{borland1995sdds} M. Borland,
\emph{A self-describing file protocol for simulation integration and shared postprocessors},
in Proceedings of the 1995 Particle Accelerator Conference (PAC'95), Dallas, TX (1996), pp.\ 2184--2186.

\bibitem{borland1998toolkit} M. Borland,
\emph{A universal postprocessing toolkit for accelerator simulation and data analysis},
in Proceedings of the 1998 International Computational Accelerator Physics Conference (ICAP'98), Monterey, CA (1998).

\bibitem{vaswani2017} A. Vaswani \emph{et al.},
\emph{Attention is all you need},
in \emph{Advances in Neural Information Processing Systems 30} (Curran Associates, 2017), p.~5998.

\bibitem{kapchinskij1959} I. M. Kapchinskij and V. V. Vladimirskij,
\emph{Limitations of proton beam current in a strong focusing linear accelerator associated with the
beam space charge}, in \emph{Proceedings of the International Conference on High Energy Accelerators and Instrumentation}
(CERN, Geneva, 1959), p.~274.

\bibitem{bassetti1980} M. Bassetti and G. A. Erskine,
\emph{Closed expression for the electrical field of a two-dimensional Gaussian charge},
CERN Report No.\ CERN-ISR-TH/80-06 (1980).

\bibitem{bane2003} K. L. F. Bane,
\emph{Short-range dipole wakefields in accelerating structures for the NLC},
SLAC Report No.\ SLAC-PUB-9663 (2003).

\bibitem{saldin1997} E. L. Saldin, E. A. Schneidmiller, and M. V. Yurkov,
\emph{On the coherent radiation of an electron bunch moving in an arc of a circle},
Nucl. Instrum. Methods Phys. Res., Sect. A \textbf{398}, 373 (1997).


\end{thebibliography}
\end{document}